\documentclass[fleqn,usenatbib]{mnras}

\usepackage{newtxtext,newtxmath}
\usepackage{color}

\usepackage[T1]{fontenc}
\usepackage{ae,aecompl}

\usepackage{graphicx}	% Including figure files
\usepackage{amsmath}	% Advanced maths commands
\usepackage{amssymb}	% Extra maths symbols

\usepackage{multirow}

\DeclareGraphicsExtensions{.png}

\newcommand{\Msolar}{\mbox{\,$\rm M_{\odot}$}}        % solar mass
\newcommand{\Lsolar}{\mbox{\,$\rm L_{\odot}$}}        % solar luminosity
  \newcommand{\Teff}{\mbox{\,\em T$_{\rm eff}$}}         % effective temperature
  \newcommand{\logTeff}{\mbox{$\log T_{\rm eff}$}}         % effective temperature
  \newcommand{\logg}{\mbox{$\log g$}}                   % surface gravity
 \newcommand{\teff}{\mbox{\,$T_{\rm eff}$}}      % Teff 
\newcommand{\lgcs}{\mbox{\,$\log g / {\rm cm\,s^{-2}}$}}        % log g
  \def\simge{\mathrel{\raise1.16pt\hbox{$>$}\kern-7.0pt
    \lower3.06pt\hbox{{$\scriptstyle \sim$}}}}           % approx ge
  \def\simle{\mathrel{\raise1.16pt\hbox{$<$}\kern-7.0pt
    \lower3.06pt\hbox{{$\scriptstyle \sim$}}}}           % approx le
\title[Pulsation in faint blue stars]{Pulsation in faint blue stars}

\author[C.~M.~Byrne \& C.~S.~Jeffery]{
Conor M. Byrne$^{1,2}$\thanks{E-mail: conor.byrne@armagh.ac.uk (CMB)}
and C. Simon Jeffery$^{1,2}$
\\
$^{1}$Armagh Observatory and Planetarium, College Hill, Armagh BT61 9DG, UK\\
$^{2}$School of Physics, Trinity College Dublin, College Green, Dublin 2, Ireland\\
}

\date{Accepted 2019 December 6. Received 2019 December 6; in original form 2019 November 6}

\pubyear{2019}

\begin{document}
\label{firstpage}
\pagerange{\pageref{firstpage}--\pageref{lastpage}}
\maketitle

\begin{abstract}
Following the discovery of blue large-amplitude pulsators (BLAPs) by the OGLE survey, additional hot, high-amplitude pulsating stars have been discovered by the Zwicky Transient Facility.  It has been proposed that all of these objects are low-mass pre-white dwarfs and that their pulsations are driven by the opacity of iron-group elements. With this expanded population of pulsating objects, it was decided to compute a sequence of post-common-envelope stellar models using the {\sc{mesa}} stellar evolution code and to examine the pulsation properties of low-mass pre-white dwarfs using non-adiabatic analysis with the {\sc{gyre}} stellar oscillation code. By including the effects of atomic diffusion and radiative levitation, it is shown that a large region of instability exists from effective temperatures of 30\,000\,K up to temperatures of at least 50\,000\,K and at a wide range of surface gravities. This encompasses both groups of pulsator observed so far, and confirms that the driving mechanism is through iron group element opacity. We make some conservative estimates about the range of periods, masses, temperatures and gravities in which further such pulsators might be observed.
\end{abstract}


%%%%%%%%%%%%%%%%%%%%%%%%%%%%%%%%%%%%%%%%%%%%%%%%%%

%%%%%%%%%%%%%%%%% BODY OF PAPER %%%%%%%%%%%%%%%%%%

\section{Introduction}

Stellar variability enables astronomers to learn more about a star or stellar system than from a star of constant brightness. For example, eclipses can be used to determine binary orbital parameters, novae can be used to learn more about close binary interactions and stars which are intrinsically unstable and pulsate can provide clues about the internal structure of a star through asteroseismology. 

The discovery of a new class of pulsating star, referred to as blue-large amplitude pulsators (BLAPs), was reported by \cite{Pietrukowicz17}. A new class of pulsating star presents challenging theoretical questions such as what the driving mechanism is, what their evolutionary status is and how many such objects could be found.  A number of stars in OGLE survey observations showed brightness variations of $0.2 - 0.4\,\rm{mag}$ over periods of $20 - 40\,$minutes. Spectroscopic follow-up on a few of these variables showed that these pulsators have an effective temperature $\teff \approx 30\,000\,\rm{K}$ and a surface gravity in the range $4.2 \le \lgcs \le 4.7$. Their evolutionary origin was uncertain, but a number of potential stellar structures was suggested. One such proposal is that BLAPs are low mass, helium-core pre-white dwarfs ($\rm{M} \approx 0.3\Msolar$) which will evolve to have the same temperature and luminosities as the observed objects \citep{Romero18,Corsico18} as they approach the white dwarf cooling track. Another is that they are helium-core burning post giants evolving toward the extended horizontal branch (EHB) \citep{Wu18,Byrne18b}.

Detailed evolutionary calculations for both pre-EHB stars and low-mass pre-white dwarfs including the effects of atomic diffusion were carried out by \cite{Byrne18b} to test the hypothesis that either model could pulsate in a manner comparable to the observations. By carrying out non-adiabatic pulsation analyses, it was shown that models of both types will evolve to become unstable in the temperature region in which BLAPs are observed, with pulsations in the appropriate period range. The driving mechanism was identified as the opacity bump  ($\kappa$-mechanism) at the iron opacity peak. \cite{Byrne18b} found that it is necessary to account for the effects of radiative levitation.

Radiative levitation is a diffusive process whereby the outgoing radiative flux from the star imparts a force on the ions in the star, with the magnitude of the force depending on the precise atomic structure of the ion. Elements such as iron have a complex `forest' of atomic transition lines and are thus more susceptible to the radiative force than relatively simple species such as helium. Stellar models including the effects of radiative levitation in hot subdwarf stars have confirmed that iron accumulates in the metal opacity bump, or Z-bump, \citep{Michaud11,Hu11} and in a self-consistent stellar evolution model, radiative levitation has been demonstrated to have a significant effect in the pre-subdwarf phase of evolution \citep{Byrne18}.

Radiative levitation enables sufficient accumulation of iron and nickel around the opacity maximum at $\sim2\times10^5\,\rm{K}$ to create a large enough opacity bump to drive the pulsations. However, it was concluded that on the basis of evolutionary timescales at the observed surface gravities, the $0.3\Msolar$ pre-white dwarf model was a more likely candidate to explain the BLAPs than the shorter-lived pre-EHB models. 

Low-mass white dwarfs ($\rm{M}\lesssim 0.5\Msolar$) are unusual in that they cannot form within the current age of the Universe through single-star evolution channels, thus requiring interaction with a binary companion to form. This can be achieved by removing mass from a red giant, either through stable mass transfer (Roche lobe overflow) or unstable mass transfer (common-envelope ejection), leaving just the helium core and a low-mass hydrogen envelope. After this interaction, the star will then contract and cool to become a white dwarf, possibly after a phase of hydrogen shell burning, depending on the mass of the hydrogen envelope that remains. Evolutionary calculations by \cite{Althaus13} have explained how these binary interactions can produce low-mass white dwarfs. Potential scenarios in which single low-mass white dwarfs could form have been explored by \cite{Kilic07} and \cite{Justham09}. 
\begin{table*}
    \centering
    \caption{Observed properties of the two groups of faint blue variable stars, alongside the rapidly-pulsating and slowly-pulsating subdwarf B stars for comparison.}
    \begin{tabular}{l|c|c|c|c}
						& `Low-gravity BLAPs'$^1$	& `High-gravity BLAPs'$^2$	& sdBVr$^3$			& sdBVs$^4$           	  \\ \hline
    Effective temperature (K)   	& $26\,000 - 32\,000$		& $31\,000 - 34\,000$		& $30\,000 - 36\,000$	& $24\,000 - 30\,000$ \\ 
    Surface gravity (\lgcs)     	& $4.20 - 4.61$				& $5.31 - 5.70$				& $5.2 - 6.1$			& $5.2 - 5.7$		  \\
    Pulsation period (s)        	& $1\,340 - 2\,360$			& $200 - 480$				& $90 - 500$			& $2\,700 - 7\,200$	  \\
    Pulsation amplitude (mag)   	& $0.2-0.4$				& $0.05-0.2$				& $0.001 - 0.064$		& $\lesssim 0.05$	  \\ \hline
    \multicolumn{5}{l}{References: 1: \cite{Pietrukowicz17}, 2: \cite{Kupfer19}, 3: \cite{Ostensen10}, 4: \cite{Green03}} \\
    \end{tabular}
    \label{tab:properties}
\end{table*}

The $\kappa$-mechanism and radiative levitation of iron and nickel are also known to be responsible for the pulsations seen in hot subdwarf stars. Hot subdwarfs are low-mass ($\sim0.46\Msolar$) core-helium-burning stars with low-mass hydrogen envelopes. Like low-mass white dwarf, hot subdwarfs also form through binary interactions removing the bulk of their hydrogen rich envelope, with the important distinction that the envelope is only stripped when the star has grown a core large enough for helium fusion to take place. There are both rapid and slow subdwarf pulsators which were discovered by \cite{Kilkenny97} and \cite{Green03} respectively. Radiative levitation of iron-group elements was predicted to be responsible for the rapid pulsators before they were discovered \citep{Charpinet96}, and the same mechanism was subsequently found to be responsible for the slow pulsators \citep{Fontaine03}. 

\begin{table}
    \centering
    \caption{Spectroscopically determined properties of BLAPs for which follow-up observations have been carried out.}
    \begin{tabular}{lcccc}
    Name			& \Teff/K		&  \lgcs	& P (s)		& Ref.	\\ \hline
    OGLE-BLAP-001	& $30\,800$	& $4.61$	& $1695.6$	& 1		\\ 
    OGLE-BLAP-009	& $31\,800$	& $4.40$	& $1916.4$	& 1		\\
    OGLE-BLAP-011	& $26\,200$	& $4.20$	& $2092.2$	& 1		\\
    OGLE-BLAP-014	& $30\,900$	& $4.42$	& $2017.2$	& 1		\\
    HG-BLAP-1		& $34\,000$	& $5.70$	& $200.20$	& 2		\\
    HG-BLAP-2		& $31\,400$	& $5.41$	& $363.16$	& 2		\\
    HG-BLAP-3		& $31\,600$	& $5.33$	& $438.83$	& 2		\\
    HG-BLAP-4		& $31\,700$	& $5.31$	& $475.48$	& 2		\\ \hline
    \multicolumn{5}{l}{Souces: 1: \cite{Pietrukowicz17}, 2: \cite{Kupfer19}} \\
    \end{tabular}
    \label{tab:BLAPs}
\end{table}

Following the discovery of BLAPs, another new group of pulsating variable stars was discovered by the Zwicky Transient Facility (ZTF) \citep{Kupfer19}. These variable stars show similar large amplitude variability as that seen in BLAPs, while their spectroscopically determined surface gravities are much higher and they have much shorter periods. These higher surface gravities and surface temperatures place them amongst the hot subdwarf pulsators in the $\logg-\logTeff$ diagram. Table~\ref{tab:properties} compares the general population properties of the \cite{Kupfer19} and \cite{Pietrukowicz17} pulsators and both classes of hot subdwarf variables, while Table~\ref{tab:BLAPs} lists the properties of the individual stars for which spectroscopic follow-up has been completed. This new discovery raises an interesting question of whether these two groups of faint, large-amplitude pulsators are related to each other and/or to the hot subdwarfs. \cite{Kupfer19} computed some helium-core pre-WD models and found that lower mass ($\sim0.28\Msolar$) pre-WDs provide a very good match in terms of the pulsation periods, while a model of a shell He-burning, post-EHB star gives pulsation periods which are longer than those observed, making this a less likely candidate. Thus the current opinion appears to be that the pulsators of \cite{Pietrukowicz17} and \cite{Kupfer19} are related objects, different only in mass. This raises the question of whether these are two distinct groups of pulsator or part of a contiguous region of instability, which may become populated as these large-scale sky surveys such as ZTF and OGLE continue to monitor the sky.

In order to investigate this, we decided to expand on the work of \cite{Byrne18b} and examine a larger region of parameter space, looking at a wide range of pre-WD masses and probe the stability of these stars as they evolve. This will enable us to determine the extent to which these pre-white dwarf stars could be unstable and make predictions about what further discoveries might be expected in this region of parameter space. 

Previous studies of pre-white dwarfs have been used to study the evolution of companions to pulsars \citep{Driebe98}. Evolutionary modelling of extremely low mass pre-WDs has been carried out by \cite{Istrate14} in the context of the timescales involved and the impact this has on the age determination of white dwarfs and also by \cite{Istrate17}, who focused on WASP 0247-25B, a pulsating pre-WD of around $0.18\Msolar\,$. As an eclipsing double-lined spectroscopic binary, WASP 0247-25B is representative of the EL\,CVn type double-lined eclipsing binaries which contain a pre-WD with an A- or F-type main-sequence companion \citep{Maxted14}.  Another group of pre-WDs includes single-lined systems containing a pre-WD with an unseen CO-WD companion \citep{Gianninas16}. Both groups may have some connection to BLAPs and both contain pulsating variables; \cite{JefferySaio13} argue that reduced hydrogen abundance in the envelope led to helium being the principle opacity driver. The evolution of EL\,CVn systems was examined by \cite{Chen17}.

\section{Methods}

A grid of models was constructed to examine a wide range of masses for pre-WD objects, from very low masses ($<0.2\Msolar$) to just below the critical mass for helium ignition, at which point the star would evolve to become a hot subdwarf prior to becoming a white dwarf ($\sim0.46\Msolar$).

\subsection{Evolution models}

Evolution models were computed using {\sc{mesa}} \citep[revision 11701]{Paxton11,Paxton13,Paxton15,Paxton18,Paxton19}. Physics options were similar to those used in \cite{Byrne18b}, namely the Schwarzchild convection criterion, a convective overshoot parameter $\alpha_{MLT}=1.9$ following \cite{Stancliffe16}, no mass loss (aside from the initial stripping of the red giant progenitor) and an initial metallicity of $Z = 0.02$ with the mixture of \cite{Grevesse98}. Atomic diffusion in {\sc{mesa}} uses the Burgers equations \citep{Burgers69}, following the approach of \cite{Thoul94}, with radiative accelerations computed using the methods outlined by \cite{Hu11}. The effects of radiative levitation make use of the monochromatic opacity data from the Opacity Project \citep{OP1,OP2} and in this work are computed for the isotopes $^{1}$H, $^{4}$He, $^{12}$C, $^{14}$N, $^{16}$O, $^{20}$Ne, $^{24}$Mg, $^{40}$Ar, $^{52}$Cr, $^{56}$Fe and $^{58}$Ni. 

An example low-mass white dwarf progenitor model was constructed starting from a $1\Msolar$ zero-age Main Sequence model. This model evolved to become a white dwarf, with a number of intermediate models saved with a range of helium core mass from $0.18\Msolar$ to $0.38\Msolar$ in steps of $0.005\Msolar$ and $0.38\Msolar$ to $0.46\Msolar$ in steps of $0.01\Msolar$. This encompasses the $0.31\Msolar$ stars found by \cite{Pietrukowicz17} and the objects of $0.28\Msolar$ which appear to reproduce well the observations of \cite{Kupfer19}. To replicate the effects of a common-envelope ejection event, each of these models was then stripped of their hydrogen envelope through a large mass-loss rate ($\dot{M} = 10^{-3}\Msolar\,\rm{yr}^{-1}$) until a relatively small envelope of approximately $3\times10^{-3}\Msolar$ remained. Each of these white dwarf progenitors was then evolved, with the effects of radiative levitation accounted for as long as the effective temperature was greater than $\sim6\,000\,\rm{K}$. Models were allowed to evolve until becoming a white dwarf with a luminosity $\log(\rm{L}/\Lsolar)=-2$. Intermediate {\sc{mesa}} models were saved every 50 time steps for pulsation analysis. Some models underwent a hydrogen shell flash, these models are discussed further in Section~\ref{sec:results}.

\begin{figure}
    \centering
    \includegraphics[width=0.47\textwidth]{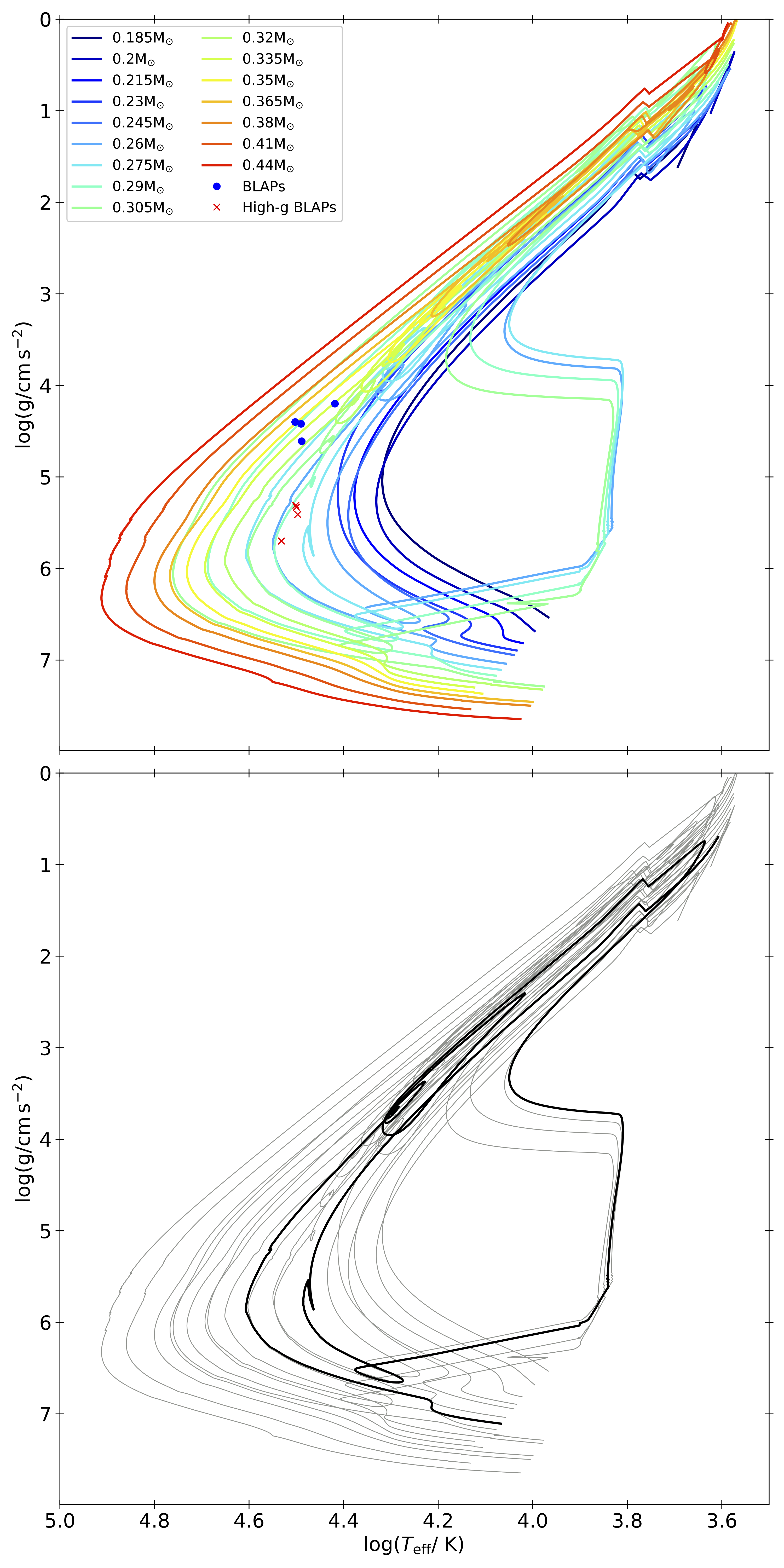}
    \caption{Illustrative evolutionary tracks in the $\logg-\logTeff$ plane of some of the models in the grid. The masses in the legend refer to the core mass. The locations of the low-gravity and high-gravity pulsators are also marked. To aid visibility of the loops in the diagram, the lower panel highlights the evolutionary track of a single model, namely the model with a core mass of $0.275\,\Msolar$ in black, with the other models in grey.}
    \label{fig:evolution}
\end{figure}

\subsection{Pulsation models}

The stellar models were analysed for stability using the {\sc{gyre}} stellar oscillation code \citep{Townsend13}. The interim {\sc{mesa}} stellar profiles from each evolution track were provided as input to {\sc{gyre}}. An adiabatic analysis is computed first to find the eigenfrequencies of the star. These adiabatic results are then used as the initial guess in a non-adiabatic analysis to determine the stability of the identified modes. As with the models analysed in \cite{Byrne18b}, the frequency scan was limited to $l = 0$ modes and focused on the fundamental radial mode given that both the prototype BLAPs and the high-gravity counterparts show large amplitude pulsations characteristic of the fundamental mode. 

Static envelope models produced by \cite{JefferySaio16} indicate that the Z-bump instability strip extends over a broader range of surface gravities than seen in hot subdwarf stars. As was the case in \cite{Byrne18b}, the blue-edge of the instability strip appears to be located at a higher temperature in the whole star evolutionary models with inhomogeneous envelope compositions than in the static homogeneous envelope calculations of \cite{JefferySaio16}. 
\section{Results}
\label{sec:results}

\begin{figure*}
    \centering
    \includegraphics[width=0.66\textwidth]{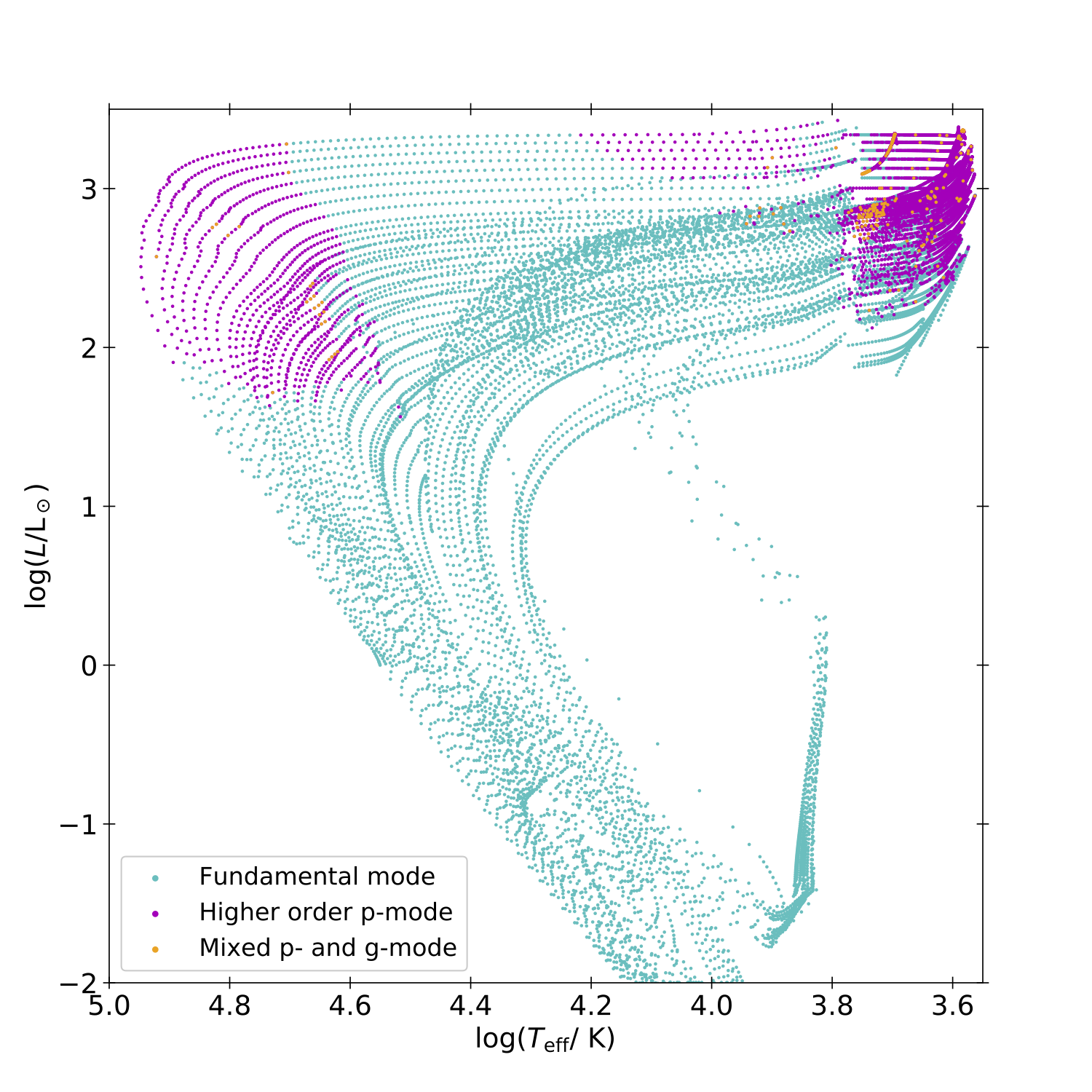}
    \caption{Lowest order mode identified by non-adiabatic {\sc{gyre}} analysis. Models where the fundamental radial mode have been identified are shown in cyan. Models where a higher order p-mode is the lowest identified are shown in magenta, while models which purport to have a g-mode component are shown in orange.}
    \label{fig:modes}
\end{figure*}

Fig~\ref{fig:evolution} shows a representative sample of the evolutionary tracks from the grid. Several behaviour characteristics are noticeable. The most massive objects evolve in a relatively simple manner, getting hotter at constant luminosity, before cooling and contracting. At the lowest masses, they likewise show a straightforward evolution. At intermediate masses ($0.255 \le \rm{M}_{\rm{core}}/\Msolar \le 0.305$) these models are observed to undergo a late hydrogen shell flash, leading them to complete a `loop' in the $\logg-\logTeff$ diagram. This ignition of helium causes expansion of the envelope, which causes the star to become a cool giant once more before returning to the white dwarf cooling sequence once more. These sorts of `late-hot flashers' are common in low mass pre-WDs and hot subdwarfs \citep{Brown01,MillerBertolami08}. Overall the evolutionary tracks are in good agreement with those shown in Fig.~1 of \cite{Istrate14}. The `kink' features in the cooling tracks of some of the models ($\logTeff\sim4.1$--4.3, $\lgcs\sim6.5$) are similar to that noted in \cite{Driebe98}, where they are identified to be models which transition to unstable hydrogen burning, but where the heat is able to escape sufficiently quickly to not go through a complete hydrogen shell flash.

\subsection{Mode identification}
\label{sec:mode_ID}

Each intermediate {\sc{mesa}} model in every evolutionary sequence was analysed non-adiabatically using {\sc{gyre}} to examine the pulsation properties of the stars as they evolve from the red giant branch to the white dwarf cooling track. The initial adiabatic analysis carried out by {\sc{gyre}} typically finds a monotonically increasing sequence of modes, starting at $n_p=1$ (the radial fundamental mode). The subsequent non-adiabatic analysis, which takes the adiabatic solutions as initial guesses to solve the non-adiabatic case generally finds the same modes of pulsation. In some situations, this is not true, with the non-adiabatic {\sc{gyre}} analysis not finding the $n=1$ mode, and in some extreme cases, the lowest order mode it identifies claims to be a p-mode mixed with a g-mode component. This is illustrated in Fig~\ref{fig:modes}, which shows the lowest order mode found by the non-adiabatic {\sc{gyre}} study. Note that this plot does not indicate the stability/instability of any model, it simply indicates the mode order of the first identified mode. The cyan symbols show the models where the fundamental mode is identified, the magenta symbols show models where a higher order radial mode is the first identified, while the orange symbols indicated models where the mode identified contains a g-mode component. There are three main regimes in which the fundamental mode is not identified. These are  $\logTeff>4.6$, $3.85\le\logTeff\le4.2$ and $3.6\le\logTeff\le3.75$, which are a result of highly non-adiabatic conditions associated with enhanced Fe opacity for the first two, and H and He opacity for the latter. The models which {\sc{gyre}} has attributed some g-mode characteristic constitute a relatively small fraction of the models and this has been attributed to numerical noise as a consequence of the highly non-adiabatic conditions in some of the models. This is a logical conclusion as radial modes, which are not related to buoyancy, should not couple to g-modes (Saio, H., priv. comm.).

\subsection{Pulsation analysis}

\begin{figure}
    \centering
    \includegraphics[width=0.49\textwidth]{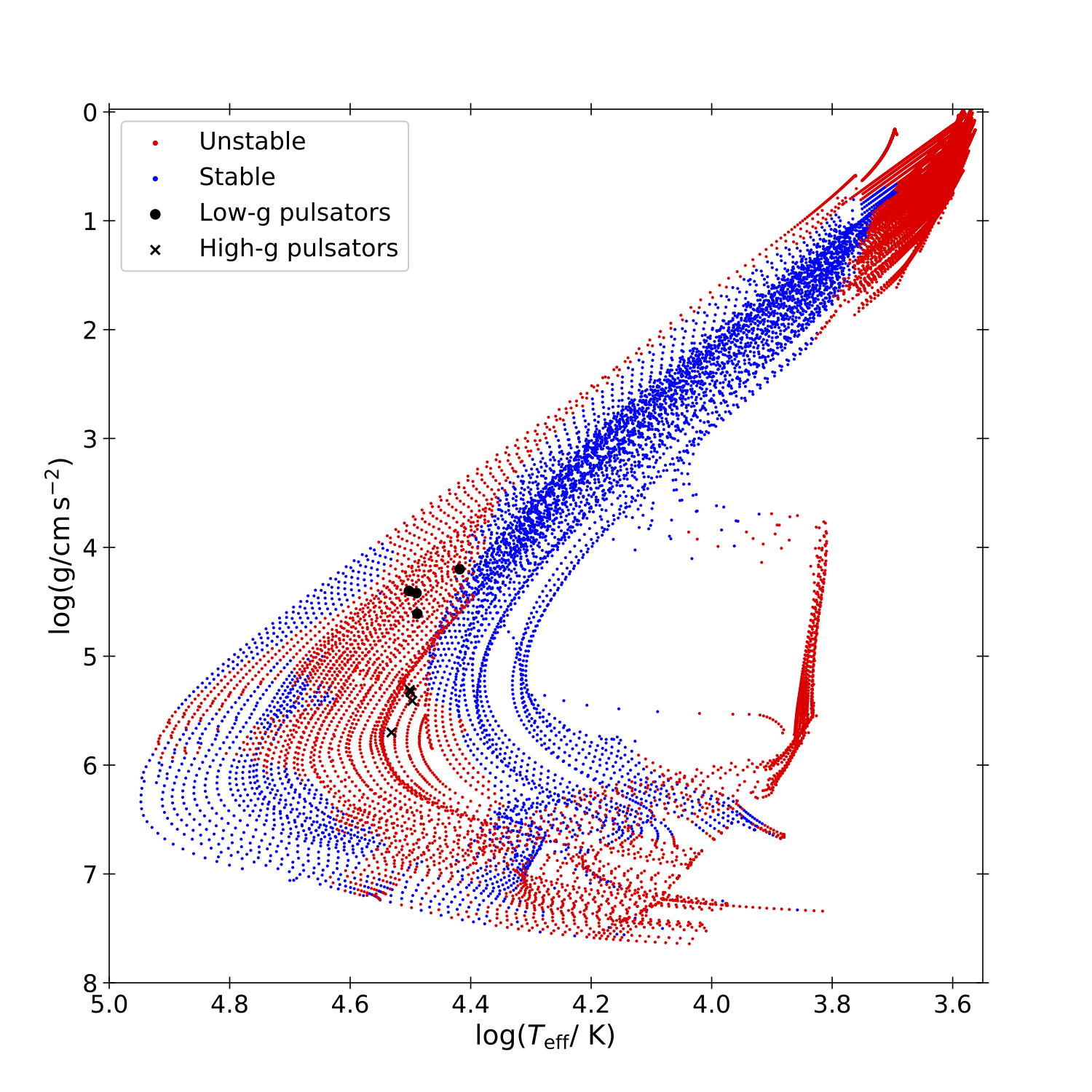}\\
     \includegraphics[width=0.49\textwidth]{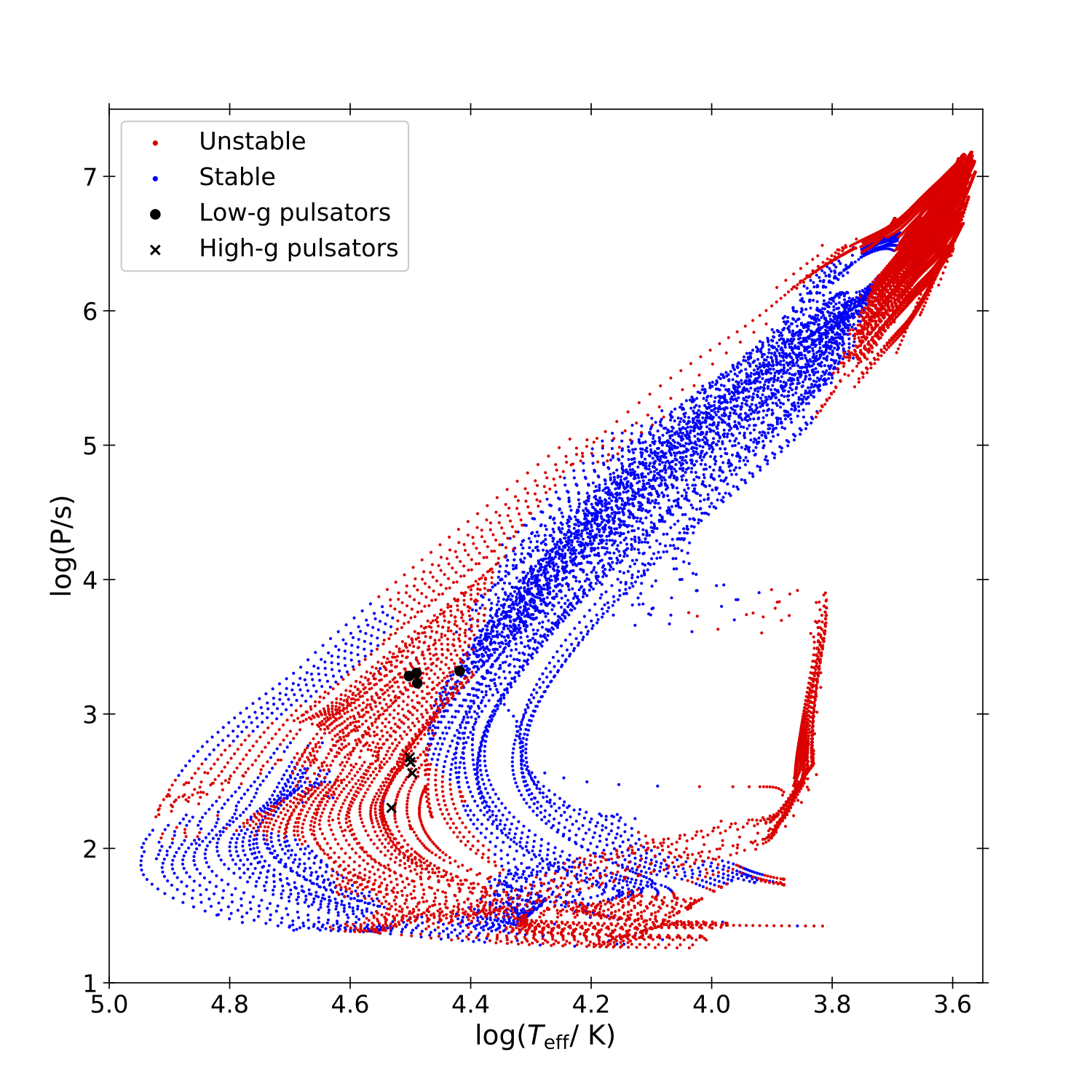}
    \caption{{\textit{Upper panel:}} The $\logg-\logTeff$ plane showing the pulsation stability of the calculated models. Models where the lowest-order mode to be identified is unstable are shown in red, while those which are stable are shown in blue. The location of the \protect\cite{Pietrukowicz17} and \protect\cite{Kupfer19} variables are also indicated. {\textit{Lower panel:}} As above, but with the logarithm of the pulsation period of the lowest order mode in seconds plotted on the y-axis.}
    \label{fig:gt_stab}
\end{figure}

 Fig~\ref{fig:gt_stab} plots each individual stellar model on the $\logg-\logTeff$ plane (upper panel) and the $\log\,P-\logTeff$ plane (lower panel) and indicates the stability of the lowest order mode identified in the frequency scan of the uniform-in-frequency grid. This grid was chosen as it is best suited to identifying p-modes. There are a number of points of interest in this diagram.

\begin{figure}
    \centering
    \includegraphics[width=0.49\textwidth]{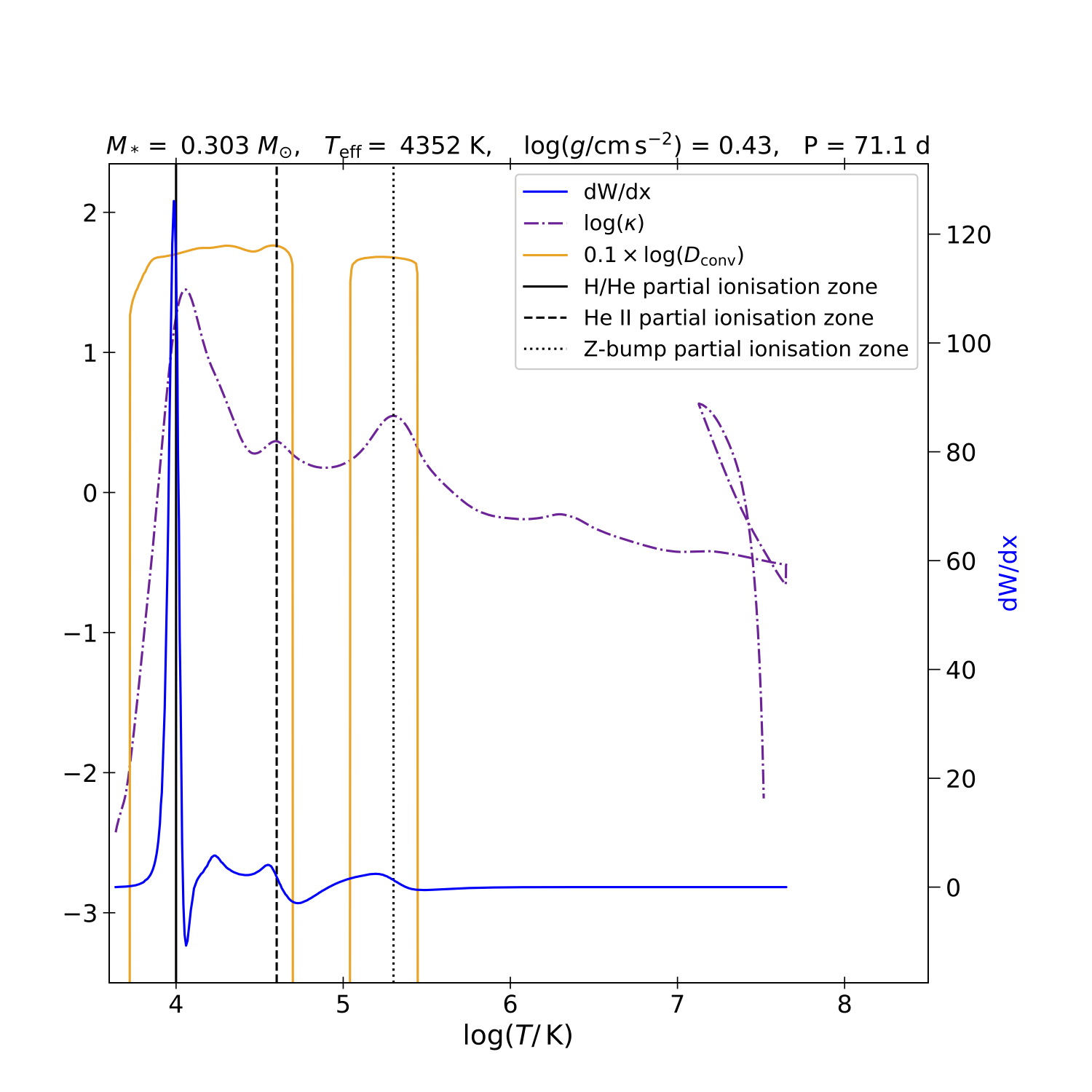}
    \caption{Profile of opacity, convective regions and work function derivative of the fundamental mode as a function of stellar interior temperature. The logarithm of the opacity is shown by dot-dashed purple line, the logarithm of the convective velocity (scaled down by a factor of 10) is indicated by the solid orange line, while the value of $\rm{d}W/\rm{d}x$ is indicated by the solid blue line. Note that the magnitude of $\rm{d}W/\rm{d}x$ is indicated on the right-hand axis, unlike the other variables. The vertical solid, dashed and dotted lines indicate the approximate temperatures of the partial ionisation opacity peaks of H/He, He{\sc{ii}} and iron group elements respectively. Note that there is a temperature inversion and hence the variables are double-valued for values of $\log T>7$.}
    \label{fig:hhe_driv}
\end{figure}

\begin{figure}
    \centering
    \includegraphics[width=0.49\textwidth]{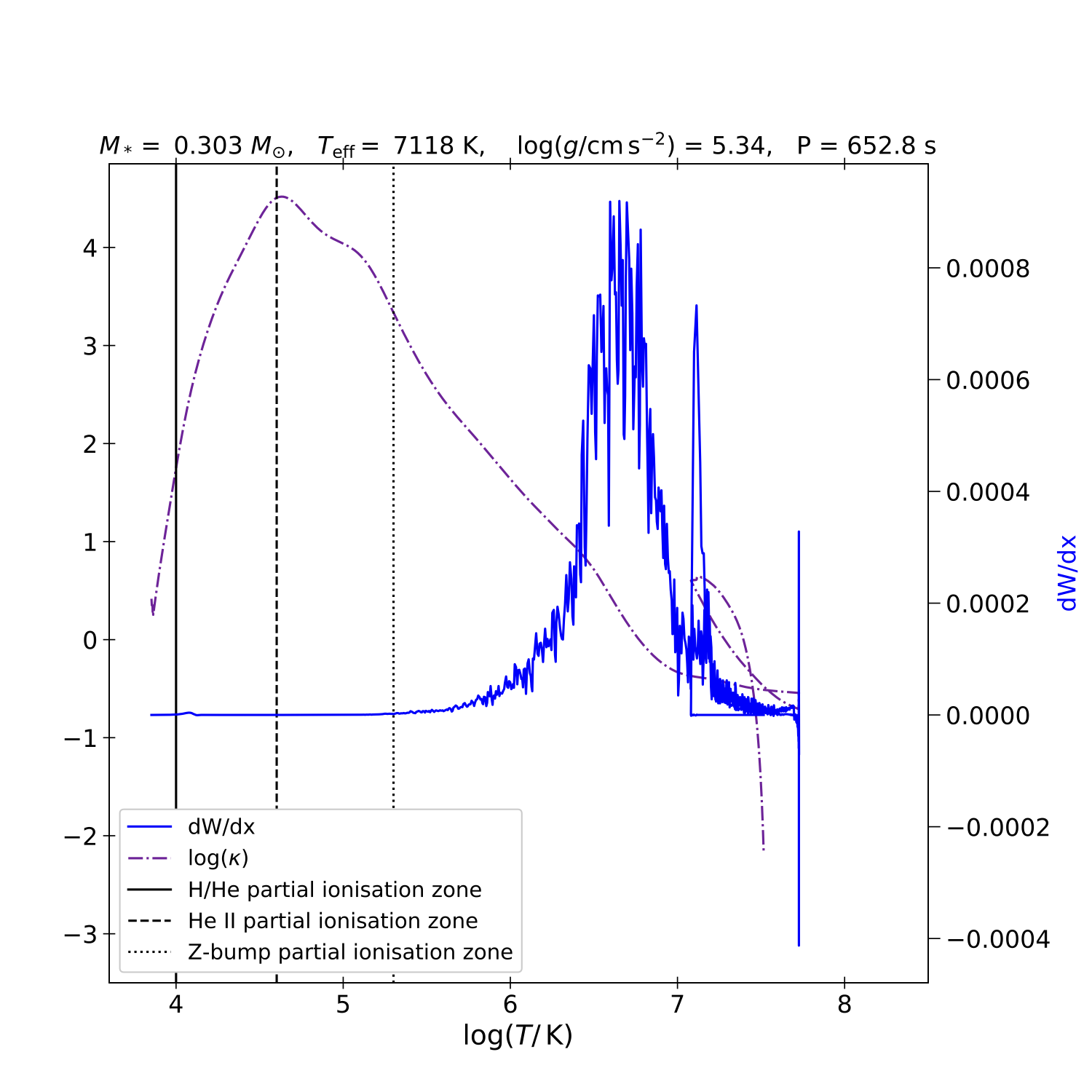}
    \caption{As in Fig~\protect\ref{fig:hhe_driv}, but for a model in the post-hydrogen shell flash evolutionary state. Again, note that there is a temperature inversion and hence the variables are double-valued for values of $\log T>7$.}
    \label{fig:driv_flash}
\end{figure}

\begin{figure}
    \centering
    \includegraphics[width=0.49\textwidth]{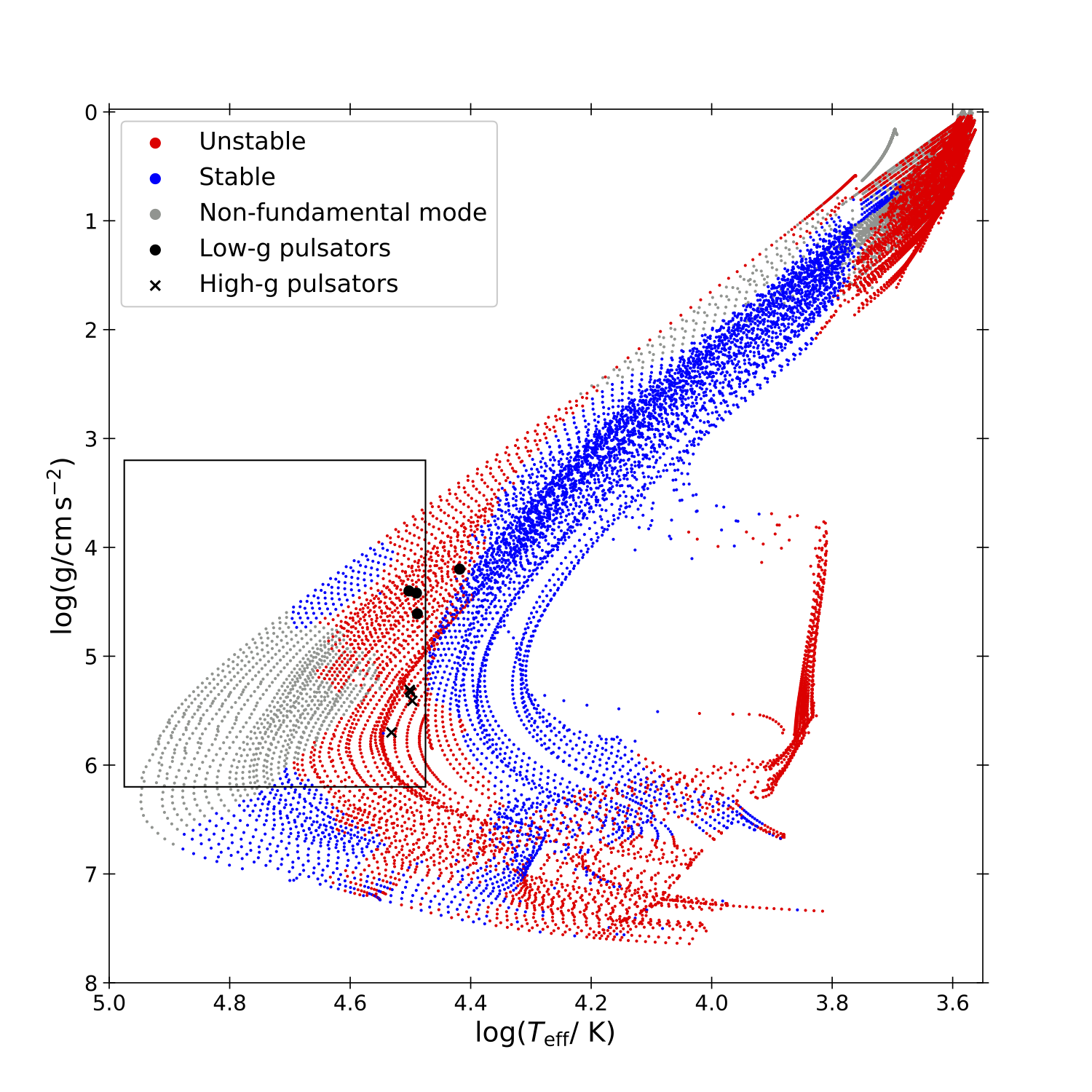}\\
    \caption{As in Fig~\protect\ref{fig:gt_stab}, except the models for which the fundamental mode was not identified are plotted in grey. The black box indicates the portion of the instability region which is discussed in more detail in Section~\ref{sec:inst_reg}.}
    \label{fig:gt_grey}
\end{figure}

\begin{itemize}
\item{At low effective temperature ($\Teff \lesssim 5000\,$K) and low surface gravity ($\lgcs \lesssim 2.0$), all the models show instability. The opacity profile of a representative model chosen from this region ($\rm{M}_{\rm{core}} = 0.3$\Msolar, $\Teff=4400\,$K, $\lgcs=0.43$) is shown in Fig~\ref{fig:hhe_driv} along with the value of $\rm{d}W/\rm{d}x$, the derivative of the work function for the fundamental mode identified by the non-adiabatic pulsation analysis. The result shows the peak in the driving is located at an effective temperature of  $10\,000\,$K, suggesting it could be the result of the H\,{\sc{i-ii}} and He\,{\sc{i-ii}} partial ionisation opacity bump, also believed to be at least partly responsible for the pulsations in Mira variables \citep{Ostlie86}. It must however be noted that the coupling between pulsation and convection is not considered by {\sc{gyre}} and therefore the nature of these pulsations cannot be deduced with certainty.}
\item{At the highest luminosities (the uppermost diagonal in the $\logg-\logTeff$ plane) the models are also unstable. This can be attributed to strange-mode pulsation due to the highly non-adiabatic nature of these stars, which have a high L/M ratio, close to the Eddington limit.}% \textcolor{red}{check this, also citation needed}}
\item{Models completing their `loop' in the diagram are found to be unstable at $\logTeff/{\rm K}<4.0$, $4.0\le\lgcs\le6.0$ as they re-ascend to the low temperature, high luminosity regime. This phase of evolution is characterised by a significant increase in radius over a relatively short period of time. An example driving/opacity diagram for one of these objects is shown in Fig~\ref{fig:driv_flash}. It can be seen that $\rm{d}W/\rm{d}x$ is extremely noisy and the magnitude of the work function is much lower than that seen in Fig~\ref{fig:hhe_driv}. The likely cause of the instability in this case is the $\epsilon$-mechanism (pulsations driven by changes in nuclear burning rates), as a consequence of the sudden short-lived increase in nuclear burning associated with the shell flash. The 0.3\Msolar core model takes 2\,000 years to evolve from the onset of the hydrogen shell flash (identifiable by the sudden change to an upward slope in the $\logg-\logTeff$ plane when the model is on the white dwarf cooling sequence) until it reaches the low temperature turning point at the top-right corner of the $\logg-\logTeff$ diagram. Less than 60 years are spent in the portion of the diagram where the star evolves from high surface gravity ($\lgcs\approx6$) to lower surface gravity ($\lgcs\approx4$) at near constant temperature ($\logTeff\approx3.85$). This makes it clear that a 1D hydrostatic model of this star is not an appropriate treatment in this phase of evolution and therefore no conclusions should be made about the stability status of the star.}
\item{Some regions of instability are noted in the white dwarf cooling tracks. These appear most prominent between $4.5\le\logTeff/{\rm K}\le4.6$ and approximately $4.1\le\logTeff/{\rm K}\le4.3$, which are in reasonable agreement with the temperature ranges of `hot-DAV' \citep{Shibahashi05} and DAV \citep{Landolt68} white dwarfs respectively. This seems like a reasonable result as all our models have hydrogen rich envelopes in this evolutionary phase as gravitational settling dominates the atomic diffusion process, leaving mostly hydrogen at the surface.}
\item{A large region of instability is present between $25\,000\,$K and $80\,000\,$K. This is the instability region caused by the iron opacity bump, and is the identified driving mechanism for at least the `low-gravity' BLAPs \citep{Byrne18b}. This instability region is the primary focus of this work.}
\item{This high effective temperature region also includes an apparent `island of stability' for surface gravities $\lgcs<6$. As discussed in Section~\ref{sec:mode_ID} above, this portion of the diagram is where {\sc{gyre}} has had issues in its non-adiabatic analysis. As a result the lowest order mode identified is not always the radial fundamental mode, but higher order p-modes. Therefore the results in this region are incomplete, as the stability of the fundamental mode remains undetermined. This is illustrated in Fig~\ref{fig:gt_grey} where the $\logg-\logTeff$ diagram is reproduced, but with the models with no identified fundamental mode plotted in grey. This makes it difficult to estimate the location of the blue-edge of the instability region, as all the high effective temperature models have inconclusive results. A black box highlights models with $\logTeff/{\rm K}>4.475$ and $\lgcs<6.2$, where the vast majority of the models appear to be unstable, which is relevant when discussing the lifetime of these objects in Section~\ref{sec:inst_reg}.}
\item{There is an overlapping region of stable and unstable models at $\logTeff\sim4.4$, $\lgcs\sim4.7$. This region where seemingly stable and unstable stars can coexist is discussed in greater detail in Section~\ref{sec:flasher}, where the models which undergo a hydrogen shell flash are discussed in more detail.}
\item{The issue with mode identification can also be illustrated in the irregular spacing of the points in the high temperature regime in the $\log\,\rm{P}-\logTeff$ plane shown in the lower panel of Fig~\ref{fig:gt_stab} which differs from the more uniform spacing found in the $\logg-\logTeff$ diagram in the upper panel. This is further evidence that the fundamental modes are not always being found as the fundamental period should scale via the period-mean density relationship, which scales with radius for a fixed mass, while the surface gravity also scales with radius, so both plots should be similar if the fundamental mode was identified each and every time.}
\end{itemize}  

\subsection{The faint blue variable instability region}
\label{sec:inst_reg}

\begin{figure*}
    \centering
    \includegraphics[width=0.99\textwidth]{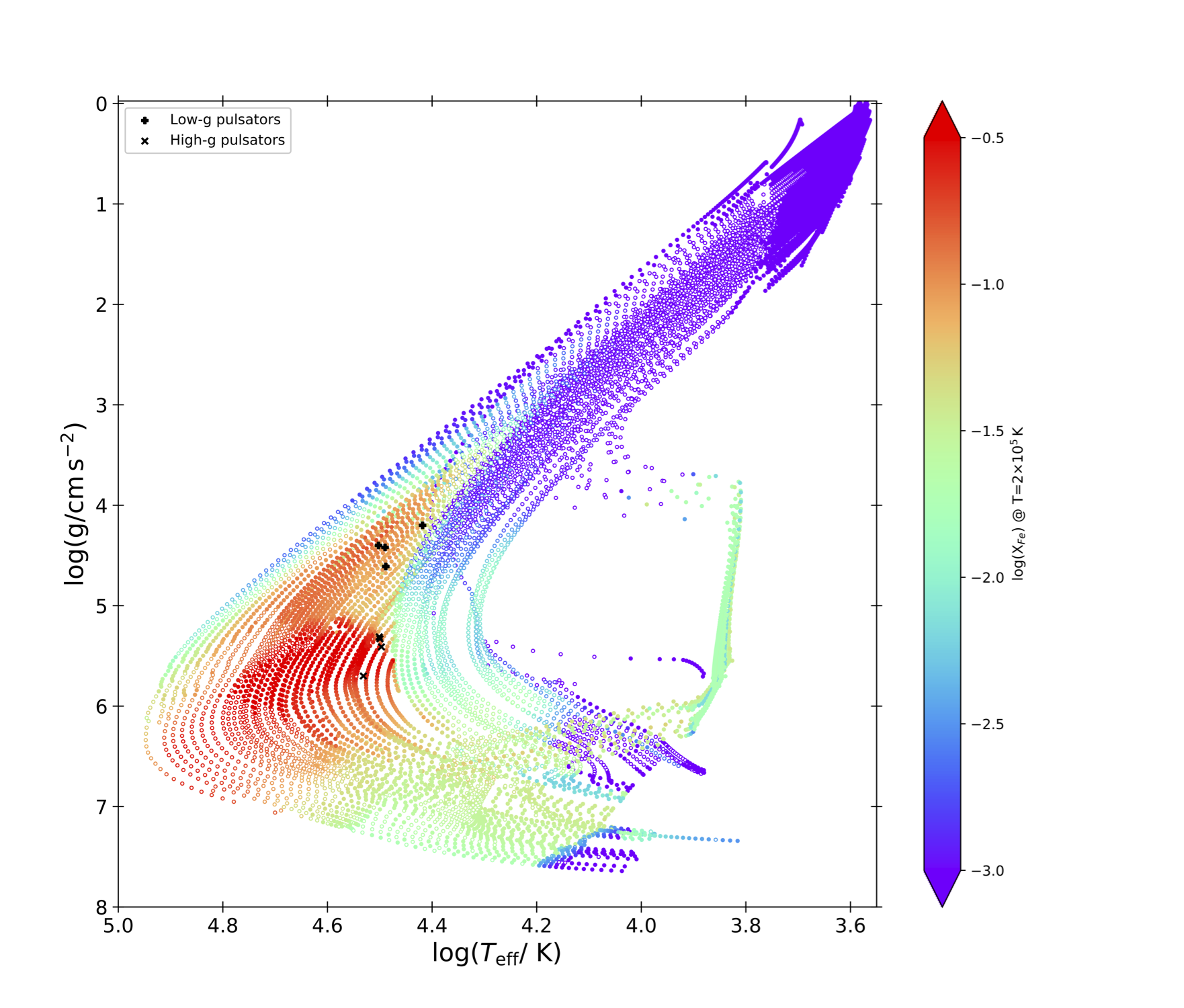}
    \caption{$\logg-\logTeff$ diagram showing the mass fraction of iron present in the iron opacity bump around T=2$\times10^5\,$K. Filled symbols indicate models with an unstable fundamental mode, while the open symbols indicate models with a stable fundamental mode. The location of the \protect\cite{Pietrukowicz17} and \protect\cite{Kupfer19} variables are also indicated.}
    \label{fig:iron}
\end{figure*}

\begin{figure}
    \centering
    \includegraphics[width=0.49\textwidth]{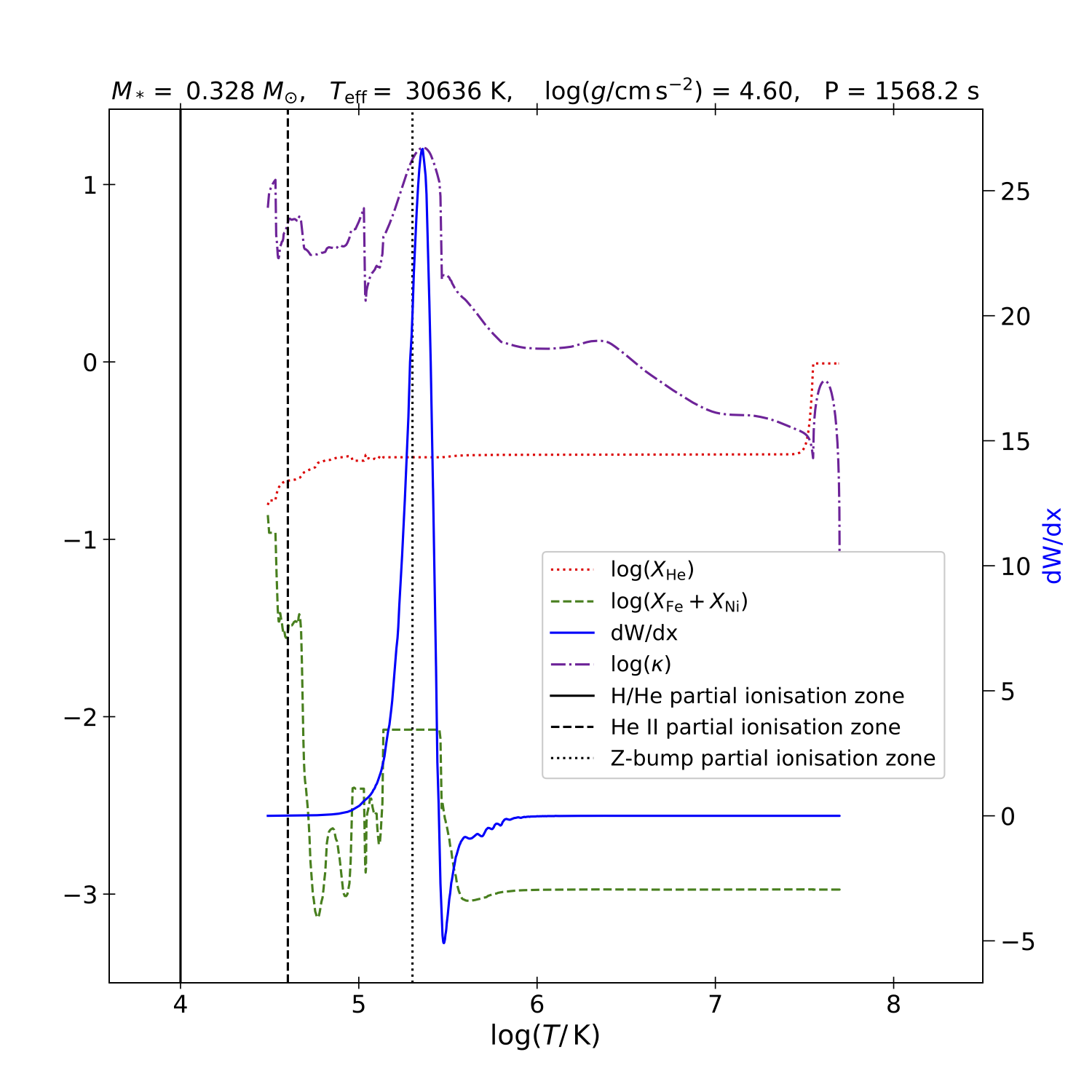}
    \includegraphics[width=0.49\textwidth]{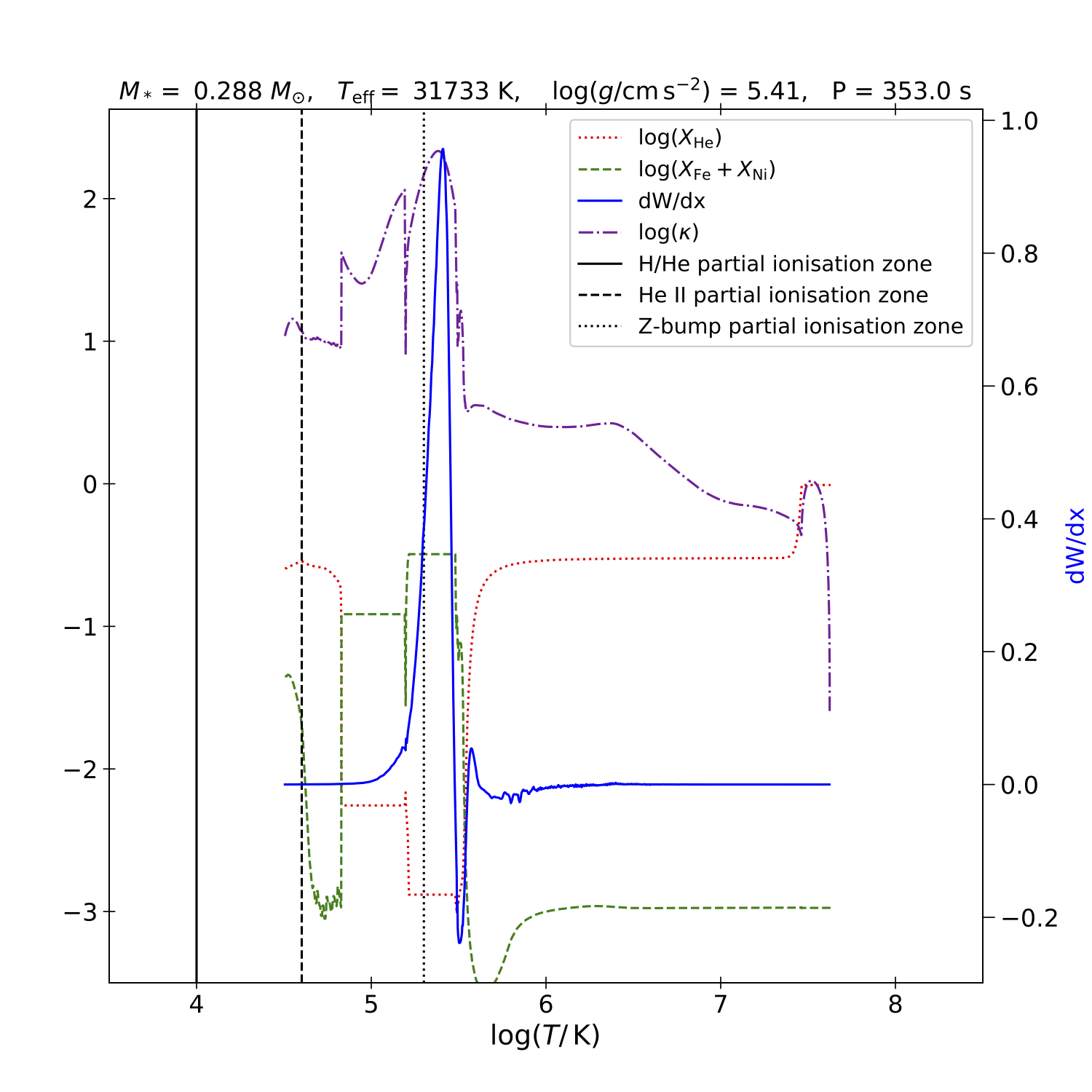}
    \caption{As in Fig~\protect\ref{fig:hhe_driv}, but for models closely resembling OGLE-BLAP-0001 (upper panel) and HG-BLAP-2  (lower panel). Additionally, the logarithm of the combined mass fraction of Fe and Ni, the mass fraction of He are shown by the dashed green and dotted red lines respectively.}
    \label{fig:blap_driv}
\end{figure}

This large instability region encompasses both the low-gravity and the high-gravity BLAPs, as well as a broader region of temperatures and gravities. Fig~\ref{fig:iron} illustrates the effects of radiative levitation on the evolution of the models. As in the previous figures, the models are placed on the $\logg-\logTeff$ diagram. In this instance, the points are colour-coded to indicate the logarithmic value of the mass fraction of $^{56}$Fe present in the region of maximum opacity ($\sim2\times10^{5}\,$K), with the purple colour at the lower end of the scale corresponding to the solar value, $\log(X_{\rm{Fe}})=-3$. The models where the lowest order mode to be identified is found to be stable are indicated by open symbols. Models where the lowest ordered mode identified is found to be unstable are indicated by the filled symbols. In general, this refers to the radial fundamental mode, except in the cases highlighted in Fig~\ref{fig:modes}, where only higher order p-modes are successfully identified. The bulk of the instability region coincides well with the iron opacity reaching its maximal value. This reinforces the result that radiative levitation of iron and nickel is the key mechanism leading to the presence of pulsations in these stars. Fig~\ref{fig:blap_driv} illustrates this for 2 example models. The upper panel shows a $\rm{M}_{\rm{core}} = 0.325\Msolar$ model with an effective temperature of $30\,600\,$K and a surface gravity of 4.6, which closely matches that of OGLE-BLAP-001 while the lower panel shows a $\rm{M}_{\rm{core}} = 0.285\Msolar$ model with an effective temperature of $31\,700\,$K and a surface gravity of 5.4, closely resembling the properties of HG-BLAP-2. In both cases, it can be seen that the peak of the driving (indicated by the maximum of the derivative of the work function) corresponds to the location of the peak in the iron abundance and corresponding opacity bump. This confirms that the driving mechanism for both the low-gravity BLAPs of \cite{Pietrukowicz17} and the high-gravity BLAPs of \cite{Kupfer19} is the {$\kappa$}-mechanism, with the opacity bump enhanced by the action of radiative levitation on iron and nickel. 

\begin{figure}
    \centering
    \includegraphics[width=0.49\textwidth]{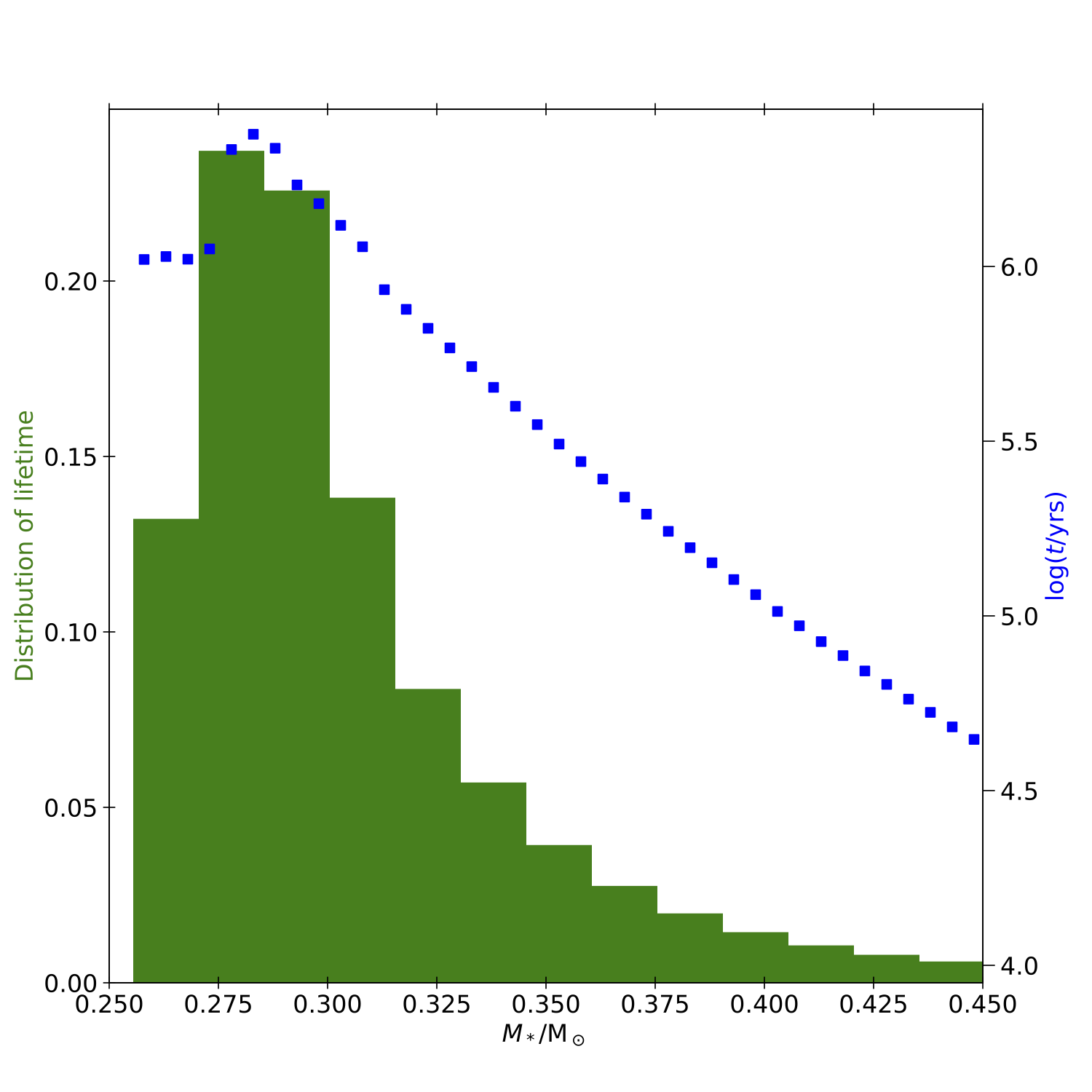}
    \caption{Lifetime of potential pulsators in the faint blue star instability region ($\logTeff\gtrsim4.475$, $\lgcs \lesssim 6.2$) as a function of mass. The blue dots indicate the logarithm of the time spent in this region. For models which undergo a hydrogen shell flash, the combined total time spent in this region during both visits to this area is the value plotted. The green histogram shows the normalised distribution of time spent in this region as a function of mass.}
    \label{fig:age}
\end{figure}

The high temperature instability region in Fig~\ref{fig:gt_stab} is a lot more extended than the region occupied by the objects which have so far been identified on the $\logg-\logTeff$ diagram. As a rough estimate, the instability region roughly extends from when the star exceeds a temperature of around $30\,000\,$K ($\logTeff=4.475$) and until it moves across at high luminosity to its maximum temperature and then until it contracts to have a surface gravity of ($\lgcs\sim6.2$). This is indicated by the rectangle in Fig~\ref{fig:gt_grey}. Considering only those models in which the fundamental mode has been identified, a majority of the models in this region are unstable. Therefore the time spent in this portion of the $\logg-\logTeff$ plane can be used as a rough indication of the lifetime of the pulsator.The stability of the models at the highest temperatures remains unclear. However, given the fact that Fig~\ref{fig:gt_stab} shows that some higher order p-modes show instability, it is not unreasonable to assume these stars are unstable for some portion of their evolution through this region. Some stars are unstable at cooler temperatures and/or higher surface gravities, but this provides a conservative estimate on where these stars could be expected to be unstable. Fig~\ref{fig:age} illustrates the time that each model spends in this portion of the $\logg-\logTeff$ plane as a function of core mass. The blue squares indicate the time spent in this region of the $\logg-\logTeff$ diagram on a logarithmic scale. The green histogram represents the mass distribution per unit time spent in the specified region of parameter space, binned into mass increments of $0.015\Msolar$. That is to say, the cumulative lifetime of the models in the area of interest is normalised such that the sum of heights of the bars in the histogram is equal to 1. This provides insight into the range of masses of objects likely to be observed in this faint blue star instability region based on the argument of evolutionary timescales, assuming a uniform proto-WD mass distribution.

This illustrates that there is an optimal range of masses in which these objects are reasonably long lived. The longest lived model in this region is the $\rm{M}_{\rm{core}} = 0.275\Msolar$ model, with models with core masses between $0.255\Msolar$ and $0.310\Msolar$ all having a `lifetime' of $10^6\,$yr, with stars with core masses greater than $0.310\Msolar$ having a logarithmically decreasing lifetime as they evolve much faster from the post-CEE phase to becoming a white dwarf. Stars with core masses below $0.255\Msolar$ don't exceed the required temperature cutoff and are not unstable in any case, so pulsating pre-WDs are not expected for masses below this. From the histogram, it can be seen that 45 per cent of the cumulative lifetime of objects is contained in the mass range $0.2705\le\rm{M}_*/\Msolar\le0.3005$, while almost 75 per cent of the cumulative lifetime is located in the mass range $0.2555\le\rm{M}_*/\Msolar\le0.3155$, with the remaining 25 per cent found at higher masses.

\subsection{Shell flash models}
\label{sec:flasher}

\begin{figure*}
    \centering
    \includegraphics[width=0.49\textwidth]{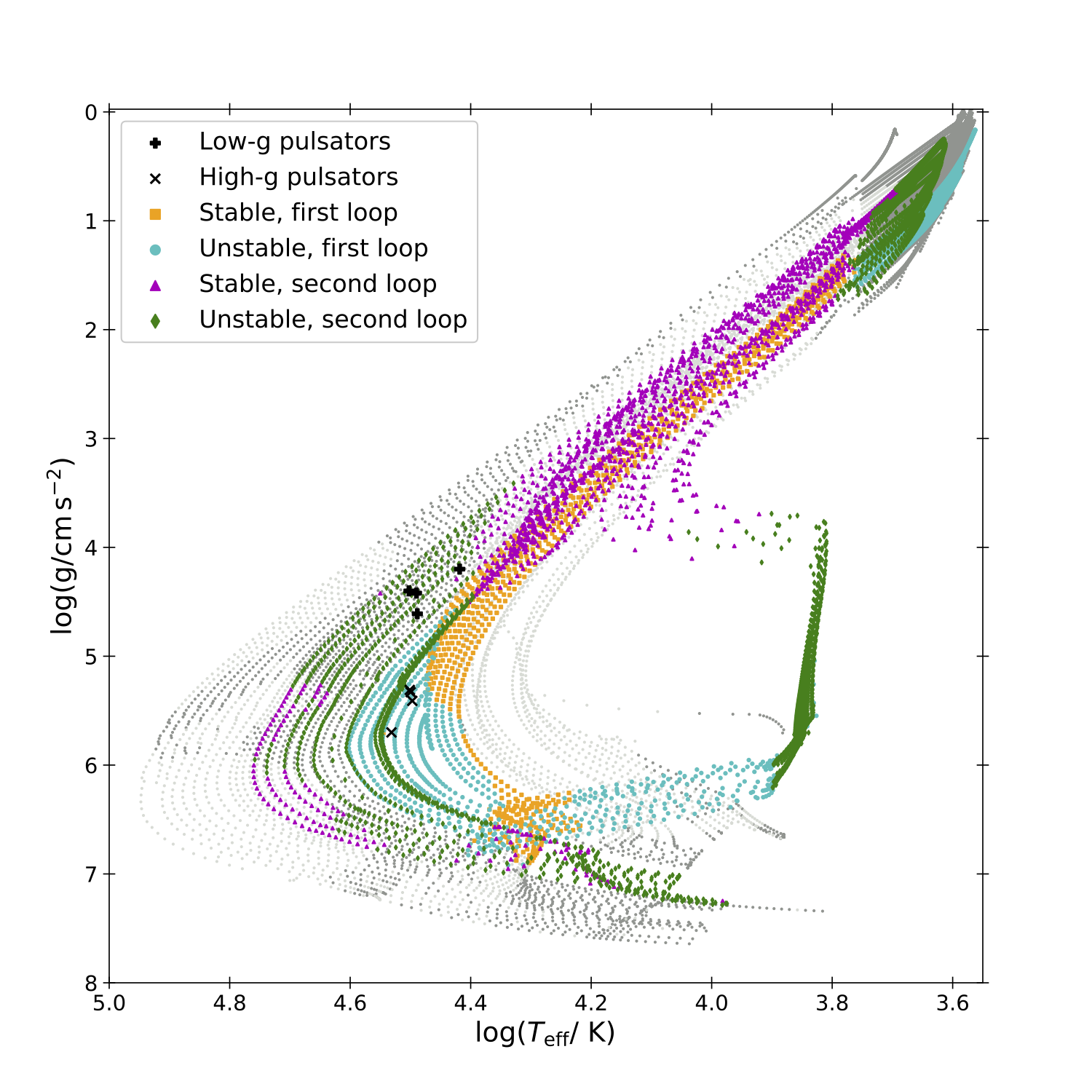}
    \includegraphics[width=0.49\textwidth]{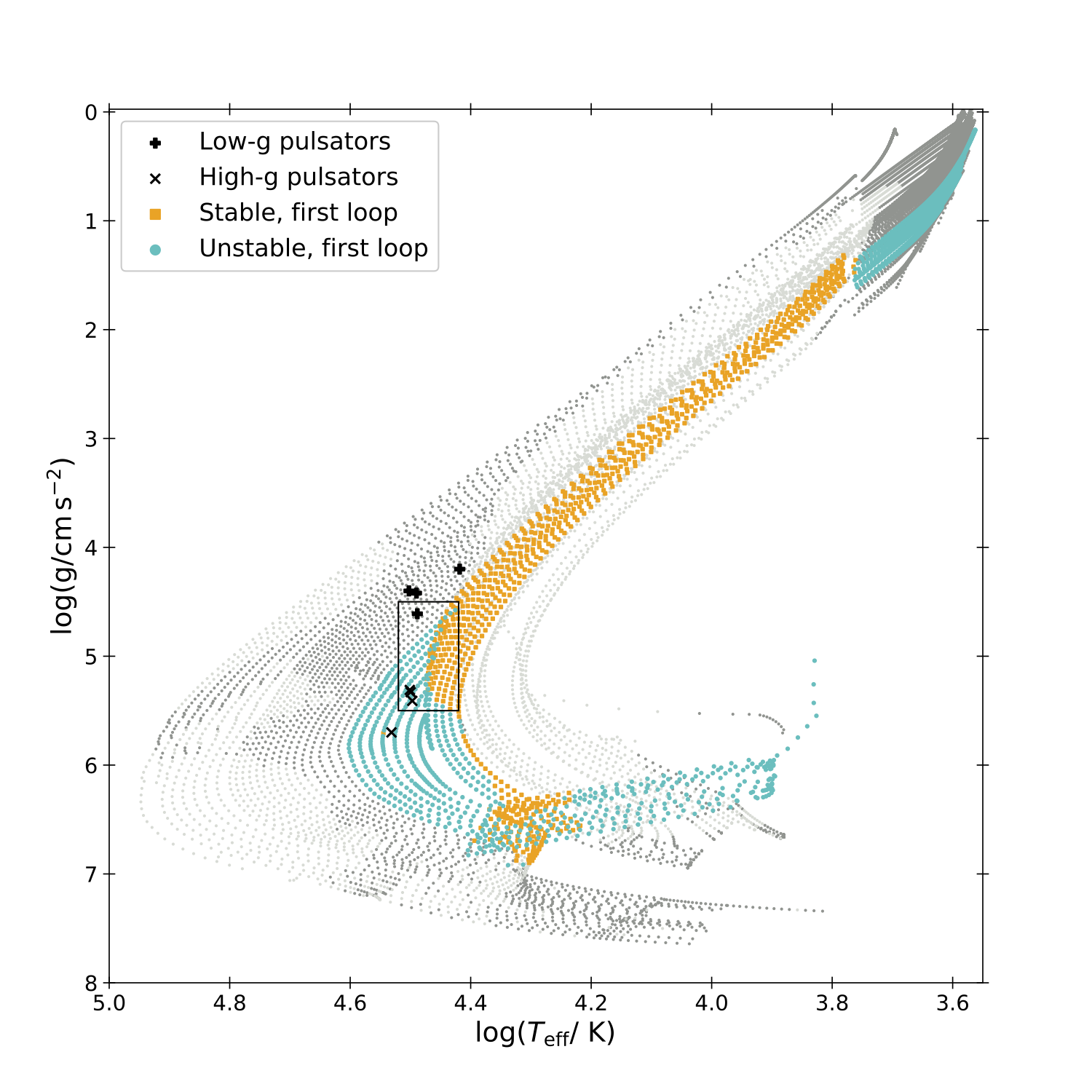}\\
    \includegraphics[width=0.49\textwidth]{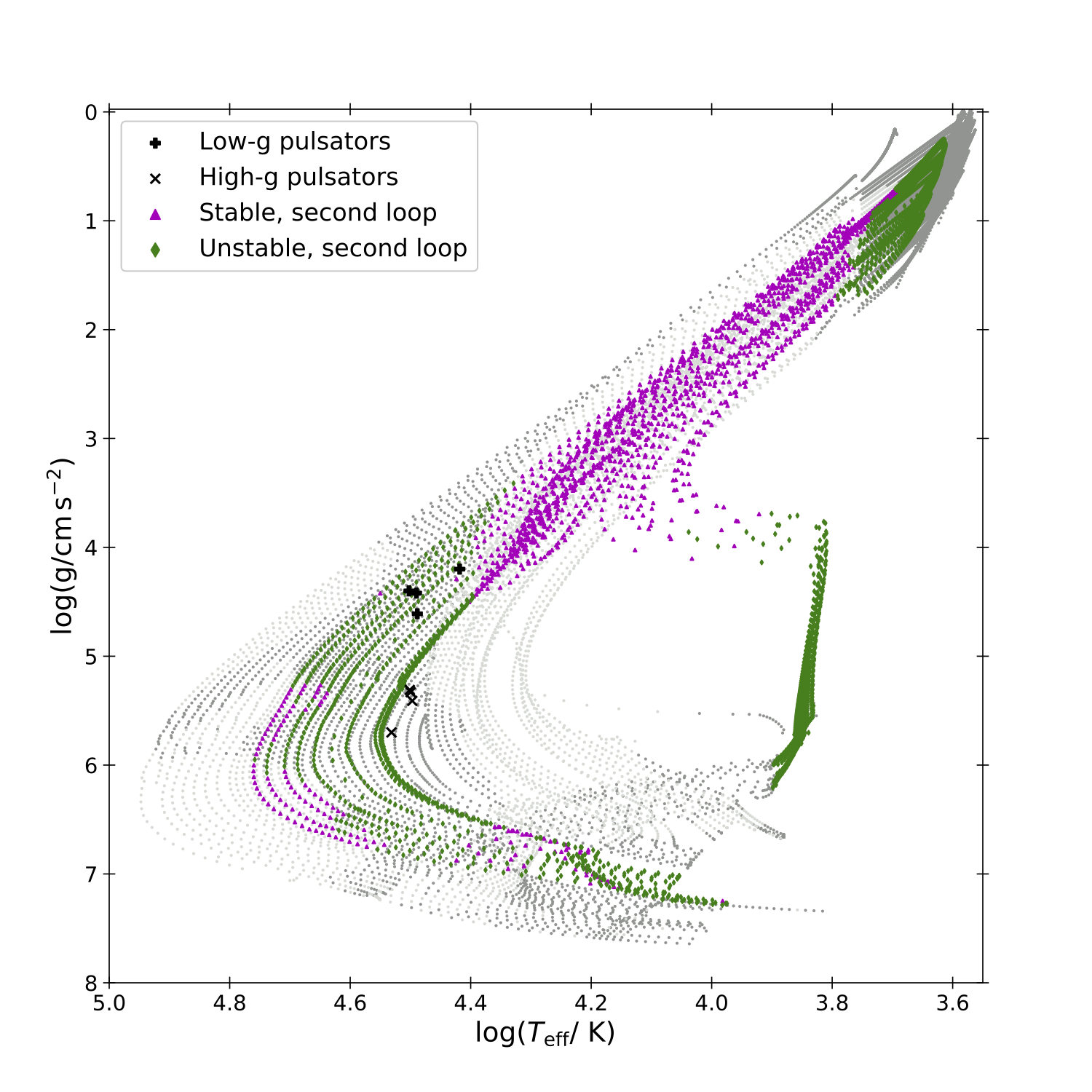}
    \includegraphics[width=0.49\textwidth]{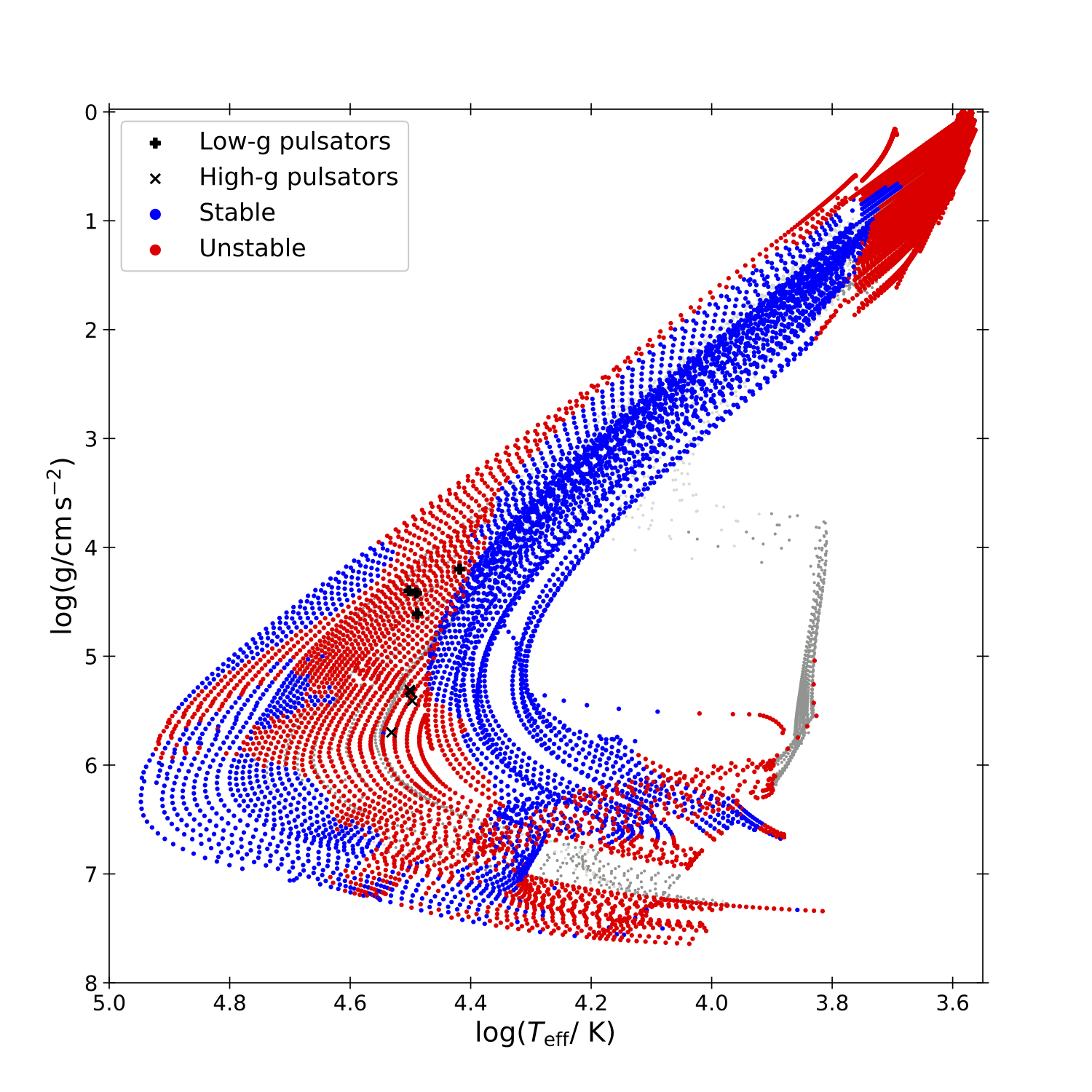}
    \caption{Stability diagram in the $\logg-\logTeff$ plane showing the difference in behaviour between the `flasher' models and the non-flashing models. {\textit{Upper left}}: Only the models with core masses between 0.255\Msolar\,and 0.305\Msolar\,which are observed to undergo a late hydrogen shell flash are coloured. The models are shown as yellow diamonds (cyan circles) when stable (unstable) on their first loop and as purple triangles (green diamonds) when stable (unstable) on their second loop. The other stable (unstable) models are shown in light grey (dark grey). {\textit{Upper right}}: Only the first loop of the `flasher' models are coloured, with the same choice of colours as in the first panel. The black rectangle indicates the area which is enlarged and shown in more detail in Fig~\protect\ref{fig:zoom_flash}. {\textit{Lower left}}: Only the second loop of the `flasher' models are coloured, with the same choice of colours as in the first panel. {\textit{Lower right}}: All models are plotted, using the same colour scheme as Fig~\protect\ref{fig:gt_stab}, but the second loop of the flasher has been left in grey. The location of the \protect\cite{Pietrukowicz17} and \protect\cite{Kupfer19} variables are also indicated as black squares and black crosses respectively in each panel.}
    \label{fig:flasher}
\end{figure*}

A number of models in this work are found to undergo a hydrogen shell flash in their post-common-envelope evolution. For the envelope mass used in this work ($3\times10^{-3}\Msolar$), this corresponds to models with a helium core mass between $0.255\Msolar\,$ and $0.305\Msolar\,$.  To examine the effects that this flash has on the pulsation stability of the models, these modes were isolated and separated into their pre-flash and post-flash loops. This is illustrated in the panels of Fig~\ref{fig:flasher}. The upper left panel shows both loops of these `flasher' models in distinct colours, with the first loop in yellow (cyan) and the second loop in magenta (green) for stable (unstable) models. The remaining models are plotted in grey, with dark grey used to indicate the unstable models. The first loop and second loop are then plotted individually in the upper right and lower left panels respectively. 

\begin{figure}
\centering
\includegraphics[width=0.49\textwidth]{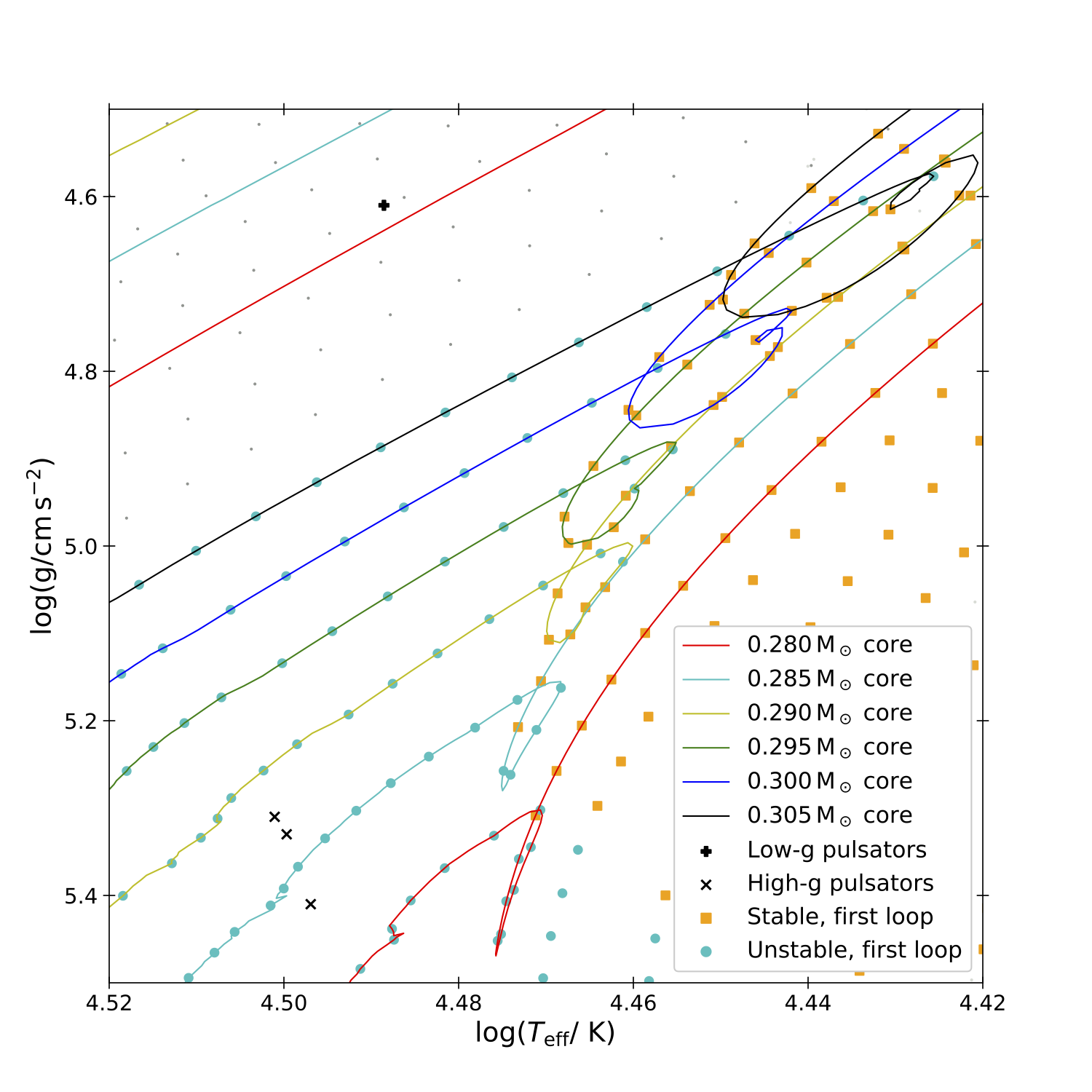}
\caption{A close-up view of the highlighted rectangle of Fig~\protect\ref{fig:flasher}, focusing on the region with overlapping stable and unstable models. To assist with clarity, the evolutionary tracks of some of the models are also drawn.}
\label{fig:zoom_flash}
\end{figure}

In the first loop, the low temperature instability region driven by the H/He partial ionisation opacity and the high temperature instability region driven by Z-bump opacity are both present and the extent of the instability is generally in agreement with the adjacent non-flasher models. A curious overlap region is observed at $\logTeff\approx4.45$ where it appears that stable and unstable models should be able to co-exist. An expanded plot of this region is shown in Fig~\ref{fig:zoom_flash}, with evolutionary tracks for clarity. In these tracks, a `hook' or `mini-loop' feature is evident, with models from before the hook/mini-loop generally being stable, and only becoming unstable while evolving through the hook/mini-loop. This hook is a result of the transition from envelope contraction following the post-common-envelope phase to the onset of a period of hydrogen shell burning once the envelope reaches the necessary conditions for shell burning to resume. The initial contraction phase is rather rapid, while the hydrogen burning phase is somewhat slower until it too is exhausted and the white dwarf contraction phase begins. For example, the $0.305\Msolar\,$ core model (shown by the black line in Fig~\ref{fig:zoom_flash}) takes $\sim10^{3}\,$yr to evolve from the post-common-envelope phase to the beginning of the loop, while taking $\sim10^5\,$yr to evolve through the loop. This rapid evolution prior to the loop is not sufficiently long relative to the diffusion timescale to allow iron and nickel to accumulate and therefore the star remains stable. Once they reach the hook/loop, the following phase of evolution is much slower, allowing the heavy metals to accumulate and pulsations are driven. The apparent overlap is due to the shape of these loops, whereby the models initially move to high effective temperature, before cooling slightly during the loop, and then becoming unstable, thus giving the appearance that some models which are slightly hotter are stable unlike their slightly cooler counterparts. Another interesting observation in this diagram is that a number of the pulsators sit quite close to this point in the evolutionary tracks, shortly after the onset of radiative levitation.

For the second loop, both of the main regions of instability, the low temperature and the high temperature regions, are in broad agreement with the rest of the overall population. The evolution of these objects after their flash is at a significantly higher luminosity and reaches a higher maximum temperature than achieved on their first loop.

\subsection{Effects of envelope mass}

\begin{figure}
\centering
\includegraphics[width=0.49\textwidth]{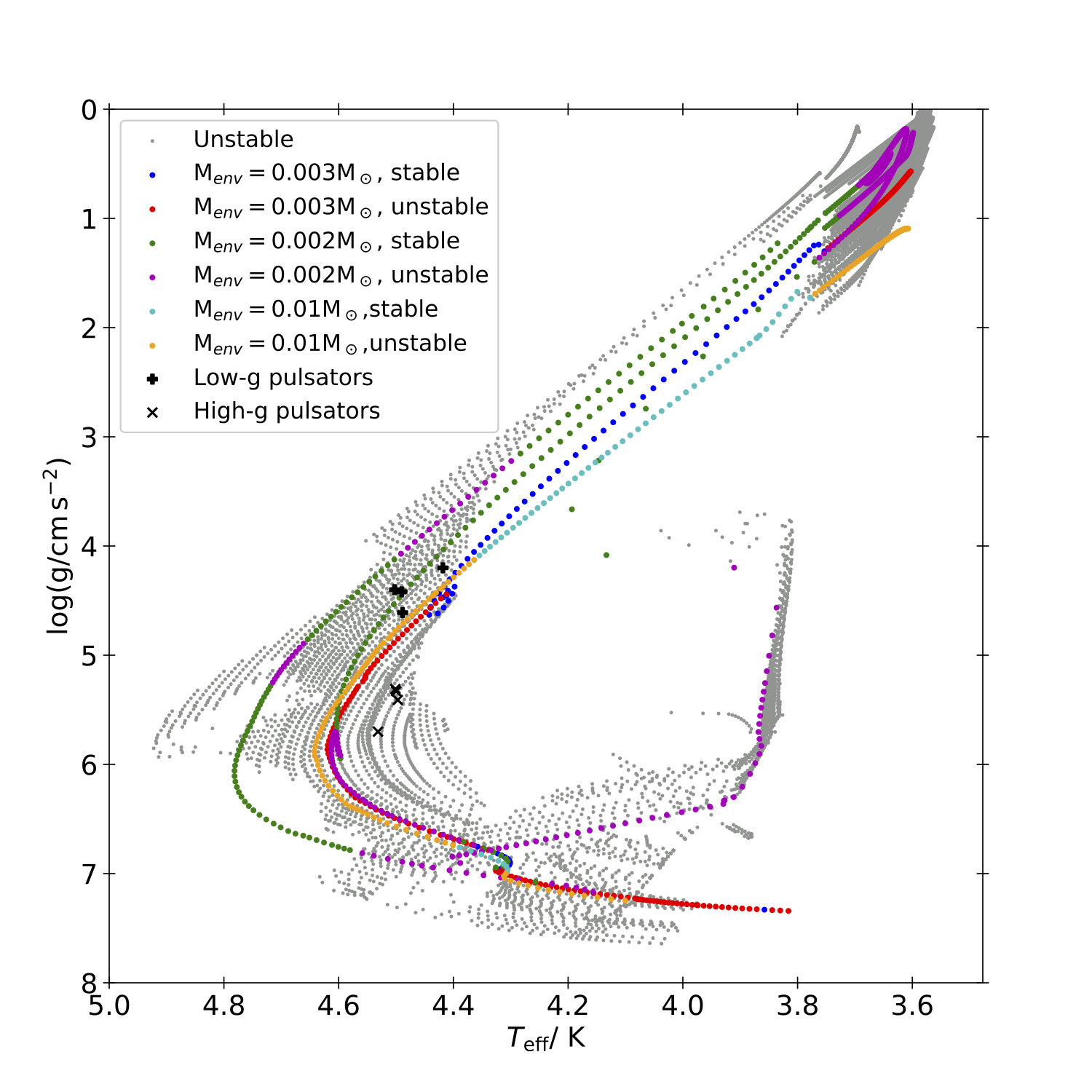}
\caption{A $\logg-\logTeff$ showing the stability of an evolutionary model with a core mass of $0.310\Msolar$ and varying envelope mass. The unstable models from the entire dataset of models with $3\times10^{-3}\Msolar\,$ envelopes from earlier figures are shown in grey for reference.}
\label{fig:envmass}
\end{figure}

The presence of shell flashes in some of the models is the result of the remnant hydrogen envelope. Therefore it is instructive to construct further models to investigate the effects of a smaller or larger hydrogen envelope on these results. This is an important area of investigation as the amount of material removed from a star in the common-envelope ejection scenario remains largely uncertain \citep[see, for example a review of this topic by][]{IvanovaRev13}. To gauge this, a smaller envelope mass model ($2\times10^{-3}\Msolar$) and a larger envelope mass model ($1\times10^{-2}\Msolar$) were computed for a $0.31\Msolar$ core model. The evolutionary tracks and the corresponding stability of the fundamental or lowest order radial mode in each model are shown in Fig~\ref{fig:envmass}.

For a smaller envelope mass, indicated by the green and magenta symbols, the model also undergoes a hydrogen shell flash. This indicates that envelope mass plays a key role in determining whether a model will have a shell flash or not, as the $3\times10^{-3}\Msolar$ envelopes primarily used in this study only lead to flashes in models with core masses up to $0.305\Msolar$. Additionally, this model has its `hook' at a much higher effective temperature than the models shown in Fig~\ref{fig:zoom_flash}, but the same principles observed in those models apply here, namely that the model remains stable during the fast contraction phase and only becomes unstable once reaching the `hook'. Apart from these differences, the instability of this model is in reasonable agreement with the instabilities of the $3\times10^{-3}\Msolar$ envelope model and the complete set of models at large. 

This significant difference in the temperature at which the hook is located is an interesting feature of the change in envelope mass, with the inference that smaller envelope masses will lead to hooks at higher temperatures. As a result, it is possible that the red edge of the instability region could provide a diagnostic by which it could determine the minimum envelope mass of a pulsator. This could help provide an indirect means of testing the physics of the common envelope phase of evolution.

For a larger envelope mass, the model evolves in a relatively similar way to the $3\times10^{-3}\Msolar$ envelope model, with a similar behaviour regarding its stability. The larger envelope mass means the model is able to sustain hydrogen shell burning in the aftermath of the common-envelope ejection, thus the star does not go through a hook/loop in its subsequent evolution. In this case, the model begins to be unstable once it reaches a high enough temperature for radiative levitation to become an effective process and has a post-common-envelope age sufficiently long compared to the diffusion timescale for iron to begin accumulating in the Z-bump.

\subsection{Other potential effects}
\label{sec:misc}

There are a number of other areas with unstable models in this data, such as in the white dwarf cooling sequence.  As the primary goal of this work was to investigate the population of high amplitude pulsators of \cite{Pietrukowicz17} and \cite{Kupfer19}, these models were not investigated in detail in this work. Additionally, as was demonstrated by changing the envelope mass, there are a number of other variables which may play a role and have an impact on the exact morphology of the instability regions. Such analysis and computation is beyond the scope of this work, but will be explored in a subsequent study. These additional effects include, but are not limited to, the method of mass loss (stable Roche lobe overflow rather than common-envelope ejection) and the initial progenitor mass. Another important factor which has not been considered in this work is the mass distribution of low-mass white dwarfs expected from population synthesis calculations. This, combined with the lifetime estimates given in Fig~\ref{fig:age} would help refine the mass range in which such objects might be detectable. It is also worth emphasising that the results presented here refer to the fundamental mode as much as was physically possible. In principle, this means that the stability of these models in higher order modes and in non-radial modes has not been examined.

\section{Discussion}

In the case of hot subdwarfs, it is known that not all stars in the appropriate temperature range are seen to be pulsators. About 10 per cent of subdwarfs in the sdBVr instability region are observed to pulsate \citep{Ostensen10}, while 75 per cent of stars in the sdbVs instability region are observed to be variable \citep{Green03}. It is unclear from this work how pure the proto-WD instability region explored in this work could be expected to be.

As this region overlaps with the hot subdwarf instability strip it is worth asking whether it is possible for objects to have been misidentified. From currently existing observations, it would appear that the amplitudes of the pre-white dwarf pulsators are considerably larger than those of the highest amplitude subdwarfs. However, moving forward it would be instructive to bear in mind that there may be two overlapping populations of objects in this part of the Hertzsprung--Russell diagram.

\subsection{Nomenclature}

As the population of these hot, subluminous variable stars grows, both in number and in extent in the $\logg-\logTeff$ plane, there is potentially some merit to discussing a coherent and descriptive naming system for these stars, as it is evident that intermediate mass/gravity objects have the potential to exist, leading to a breakdown in the dichotomy of `high-gravity' and `low-gravity' BLAPs. While their temperatures and brightnesses place them close to hot subdwarf variables, these stars are fundamentally different (shell-hydrogen burning, rather than the hot subdwarfs which are core-helium burning objects). Another potential naming convention would be to describe them as `Faint Blue Variables', as an analogue to luminous blue variables, which also show dramatic changes in brightness. Given that the variability of the luminous blue variables is an unpredictable rather than periodic variations, this name would likely cause confusion also. What is evident at this point is that all of these stars are blue, subluminous and exhibit radial mode pulsations. Perhaps a name such as faint-blue radial pulsators would be a more appropriate descriptor of this new and growing class of pulsating star.

\subsection{Completeness of population}

With the growing number of BLAPs being discovered, it becomes crucial to carry out spectroscopic analysis of as many such stars as possible So far, only 4 of the 14 OGLE pulsators have had effective temperatures and surface gravities determined spectroscopically. Locating these objects on the $\logg-\logTeff$ diagram and comparing to the stability diagrams produced in this work will enable a broader understanding of this new class of pulsator. With continuing and new high-cadence photometric surveys, it is inevitable that more such stars will be discovered in the future. These must be combined with distances \citep[e.g.][]{Ramsay18} to construct volume-limited samples and hence to test stellar evolution and binary population synthesis models for pre-white dwarfs.

\subsection{Relation to EL\,CVn stars}

From the orbital parameters of the currently known EL\,CVn systems, \cite{Chen17} show that the formation of the low-mass pre-white dwarfs in those systems cannot come from the common envelope channel as systems with a red giant with such a low core mass would lead to a merger rather than a common envelope and subsequent binary system. As discussed in Section~\ref{sec:misc} above, the Roche lobe overflow evolution channel and the effects it may have on the instability regions have not been explored in this work and remain an area for future study. 

\subsection{Binarity}

Following the implication that these pre-white dwarf variables are products of binary evolution, it would be expected that some, if not all, of the observed pulsators should show evidence of binarity. Currently none have shown any such evidence and further observations are needed to test this expectation. If any of the observed pulsators are descendants of EL\,CVn stars,  an A- or F-type main-sequence star companion should be visible in the spectrum and at least some should still be eclipsing, despite the smaller radius of the contracting pre-white dwarf. No signatures of eclipse have been found. In the case of post-common-envelope pre-white dwarfs with compact companions such as the \cite{Gianninas16} pulsators, evidence should be sought from radial-velocity studies extending over both the pulsation and putative orbital periods.  

\subsection{Extent of the instability region}

The blue-edge of the instability region remains uncertain owing to the current lack of data on the fundamental mode in our models. For the assumed envelope mass ($3\times10^{-3}\Msolar$), the red-edge is reasonably well defined, with almost all models above 30\,000\,K showing instability, with the red-edge dropping to about 25\,000\,K for the models with the highest and the lowest surface gravities. From the lower panel in Fig~\ref{fig:gt_stab}, this corresponds to pulsation periods from $\sim100\,$s as the models approach the white dwarf cooling sequence up to $\sim10\,000\,$s ($2.8\,$hr) for the more massive models as they contract at high luminosity. The red-edge is sensitive to the assumed envelope mass, and this should be explored in more detail in future.

\section{Conclusions}

Evolutionary models of low mass pre-white dwarfs including the effects of radiative levitation have provided tremendous insight into the regime in which a new and growing population of faint blue pulsating star has been discovered. A large region of the $\lgcs-\logTeff$ diagram shows instability of the radial fundamental mode from effective temperatures  around 30\,000\,K, up to at least 50\,000\,K and potentially up to 80\,000\,K. In this temperature range, the periods of the unstable fundamental modes range from around 100 seconds up to 2-3 hours at the longest durations.

The driving mechanism is clearly identified as the $\kappa$-mechanism as a result of a large Z-bump opacity as a direct consequence of radiative levitation, with the significant enhancement of heavy metal opacity providing a strong driving force. This region of instability includes both the low-gravity variables discovered by \cite{Pietrukowicz17} and their high-gravity counterparts discovered by \cite{Kupfer19}. This confirms that both of these groups of stars are part of the same phenomenon, that is that they are all pulsating pre-low mass white dwarf stars. The location of the instability region coincides remarkably well with the onset of radiative levitation, illustrating the importance of the increase in iron abundance in the Z-bump opacity peak for pulsation driving.

Assuming a flat initial mass function for pre-WDs, over 75 per cent of the cumulative time within the instability zone is spent by stars with mass $0.2555\le\rm{M}_*/\Msolar\le0.3155$. Higher mass models evolve faster. Models with masses less than $0.255\Msolar$ do not reach high enough effective temperatures for radiative levitation to become efficient enough to lead to pulsations. Models with core masses between 0.255\Msolar\, and 0.305\Msolar\, undergo a hydrogen shell flash. Closer examination of these models shows that their post-common envelope evolution is characterised by a quick increase in effective temperature as their inert hydrogen envelope remnant contracts, followed by a hook/loop as hydrogen-shell burning resumes. This leads to evolution on a nuclear timescale which allows enough time for radiative levitation to take effect and make the stars become unstable. A number of the observed pulsators are located close to this region at the red-edge of the instability region.

Experiments with differing envelope masses showed that in general the shape and extent of the instability region is largely unchanged. At larger envelope masses, the star is more likely to maintain hydrogen shell burning after common-envelope ejection and thus does not show a hook/loop feature when hydrogen re-ignites. For lower envelope masses it was found that the effective temperature of the model is much higher when it resumes hydrogen shell burning, thus shifting the effective red-edge of the instability region to higher temperatures. This suggests that with a larger population of such pulsators, the location of the red-edge may provide a diagnostic for determining the minimum envelope mass of these pre-white dwarfs, thus providing an indirect way of measuring the efficiency of common-envelope ejection at removing mass in close binary situations. 

A number of uncertainties remain in our understanding of this new class of pulsating star. The calculation of non-adiabatic eigenfrequencies with {\sc{gyre}} becomes challenging in extremely non-adiabatic models with high L/M ratios and/or large opacity bumps. It is hoped that future development will enable more robust non-adiabatic analysis with {\sc{gyre}} and provide a complete picture of the instability region at the highest effective temperatures.  This paper successfully interprets BLAPs as likely low-mass pre-white dwarfs. Such stars must be produced from a close binary interaction which strips most of the hydrogen envelope. It therefore remains puzzling that no evidence for a binary companions to any of these pulsators has yet been found, although it must be acknowledged that systematic searches for binarity (through radial velocity variations or other means) have not yet been carried out for either the low-gravity or high-gravity pulsators.

We propose that given that both the high-gravity and low-gravity stars under discussion are probably compact  fundamental radial-mode pulsators driven by the same pulsation mechanism and are the same kind of stellar object,  it may be worth providing the class of stars with a name that better reflects their properties, such as faint-blue radial pulsators.

\section*{Acknowledgements}
The authors thank Hideyuki Saio for useful discussion which have improved the content of this manuscript.
CMB acknowledges funding from the Irish Research Council's Government of Ireland Postgraduate Scholarship scheme (Grant No. GOIP 2015/1603).
The Armagh Observatory and Planetarium is funded by direct grant from the Northern Ireland Department for Communities.
The authors thank the referee for their constructive comments on this manuscript.

%%%%%%%%%%%%%%%%%%%%%%%%%%%%%%%%%%%%%%%%%%%%%%%%%%



\bibliographystyle{mnras}
\bibliography{references}

\begin{thebibliography}{}
\makeatletter
\relax
\def\mn@urlcharsother{\let\do\@makeother \do\$\do\&\do\#\do\^\do\_\do\%\do\~}
\def\mn@doi{\begingroup\mn@urlcharsother \@ifnextchar [ {\mn@doi@}
  {\mn@doi@[]}}
\def\mn@doi@[#1]#2{\def\@tempa{#1}\ifx\@tempa\@empty \href
  {http://dx.doi.org/#2} {doi:#2}\else \href {http://dx.doi.org/#2} {#1}\fi
  \endgroup}
\def\mn@eprint#1#2{\mn@eprint@#1:#2::\@nil}
\def\mn@eprint@arXiv#1{\href {http://arxiv.org/abs/#1} {{\tt arXiv:#1}}}
\def\mn@eprint@dblp#1{\href {http://dblp.uni-trier.de/rec/bibtex/#1.xml}
  {dblp:#1}}
\def\mn@eprint@#1:#2:#3:#4\@nil{\def\@tempa {#1}\def\@tempb {#2}\def\@tempc
  {#3}\ifx \@tempc \@empty \let \@tempc \@tempb \let \@tempb \@tempa \fi \ifx
  \@tempb \@empty \def\@tempb {arXiv}\fi \@ifundefined
  {mn@eprint@\@tempb}{\@tempb:\@tempc}{\expandafter \expandafter \csname
  mn@eprint@\@tempb\endcsname \expandafter{\@tempc}}}

\bibitem[\protect\citeauthoryear{{Althaus}, {Miller Bertolami}  \&
  {C{\'o}rsico}}{{Althaus} et~al.}{2013}]{Althaus13}
{Althaus} L.~G.,  {Miller Bertolami} M.~M.,   {C{\'o}rsico} A.~H.,  2013,
  \mn@doi [\aap] {10.1051/0004-6361/201321868}, \href
  {https://ui.adsabs.harvard.edu/abs/2013A&A...557A..19A} {557, A19}

\bibitem[\protect\citeauthoryear{{Brown}, {Sweigart}, {Lanz}, {Landsman}  \&
  {Hubeny}}{{Brown} et~al.}{2001}]{Brown01}
{Brown} T.~M.,  {Sweigart} A.~V.,  {Lanz} T.,  {Landsman} W.~B.,   {Hubeny} I.,
   2001, \mn@doi [\apj] {10.1086/323862}, \href
  {http://adsabs.harvard.edu/abs/2001ApJ...562..368B} {562, 368}

\bibitem[\protect\citeauthoryear{{Burgers}}{{Burgers}}{1969}]{Burgers69}
{Burgers} J.~M.,  1969, {Flow Equations for Composite Gases}.
Academic Press

\bibitem[\protect\citeauthoryear{{Byrne} \& {Jeffery}}{{Byrne} \&
  {Jeffery}}{2018}]{Byrne18b}
{Byrne} C.~M.,  {Jeffery} C.~S.,  2018, \mn@doi [\mnras]
  {10.1093/mnras/sty2545}, \href
  {https://ui.adsabs.harvard.edu/abs/2018MNRAS.481.3810B} {481, 3810}

\bibitem[\protect\citeauthoryear{{Byrne}, {Jeffery}, {Tout}  \& {Hu}}{{Byrne}
  et~al.}{2018}]{Byrne18}
{Byrne} C.~M.,  {Jeffery} C.~S.,  {Tout} C.~A.,   {Hu} H.,  2018, \mn@doi
  [\mnras] {10.1093/mnras/sty158}, \href
  {http://adsabs.harvard.edu/abs/2018MNRAS.tmp..162B} {475, 4728}

\bibitem[\protect\citeauthoryear{{Charpinet}, {Fontaine}, {Brassard}  \&
  {Dorman}}{{Charpinet} et~al.}{1996}]{Charpinet96}
{Charpinet} S.,  {Fontaine} G.,  {Brassard} P.,   {Dorman} B.,  1996, \mn@doi
  [\apjl] {10.1086/310335}, \href
  {http://adsabs.harvard.edu/abs/1996ApJ...471L.103C} {471, L103}

\bibitem[\protect\citeauthoryear{{Chen}, {Maxted}, {Li}  \& {Han}}{{Chen}
  et~al.}{2017}]{Chen17}
{Chen} X.,  {Maxted} P.~F.~L.,  {Li} J.,   {Han} Z.,  2017, \mn@doi [\mnras]
  {10.1093/mnras/stx115}, \href
  {https://ui.adsabs.harvard.edu/abs/2017MNRAS.467.1874C} {467, 1874}

\bibitem[\protect\citeauthoryear{{C{\'o}rsico}, {Romero}, {Althaus}, {Pelisoli}
   \& {Kepler}}{{C{\'o}rsico} et~al.}{2018}]{Corsico18}
{C{\'o}rsico} A.~H.,  {Romero} A.~D.,  {Althaus} L. r.~G.,  {Pelisoli} I.,
  {Kepler} S.~O.,  2018, arXiv e-prints, \href
  {https://ui.adsabs.harvard.edu/abs/2018arXiv180907451C} {p. arXiv:1809.07451}

\bibitem[\protect\citeauthoryear{{Driebe}, {Schoenberner}, {Bloecker}  \&
  {Herwig}}{{Driebe} et~al.}{1998}]{Driebe98}
{Driebe} T.,  {Schoenberner} D.,  {Bloecker} T.,   {Herwig} F.,  1998, \aap,
  \href {https://ui.adsabs.harvard.edu/abs/1998A&A...339..123D} {339, 123}

\bibitem[\protect\citeauthoryear{{Fontaine}, {Brassard}, {Charpinet}, {Green},
  {Chayer}, {Bill{\`e}res}  \& {Randall}}{{Fontaine} et~al.}{2003}]{Fontaine03}
{Fontaine} G.,  {Brassard} P.,  {Charpinet} S.,  {Green} E.~M.,  {Chayer} P.,
  {Bill{\`e}res} M.,   {Randall} S.~K.,  2003, \mn@doi [\apj] {10.1086/378270},
  \href {http://adsabs.harvard.edu/abs/2003ApJ...597..518F} {597, 518}

\bibitem[\protect\citeauthoryear{{Gianninas}, {Curd}, {Fontaine}, {Brown}  \&
  {Kilic}}{{Gianninas} et~al.}{2016}]{Gianninas16}
{Gianninas} A.,  {Curd} B.,  {Fontaine} G.,  {Brown} W.~R.,   {Kilic} M.,
  2016, \mn@doi [\apjl] {10.3847/2041-8205/822/2/L27}, \href
  {https://ui.adsabs.harvard.edu/abs/2016ApJ...822L..27G} {822, L27}

\bibitem[\protect\citeauthoryear{{Green} et~al.,}{{Green}
  et~al.}{2003}]{Green03}
{Green} E.~M.,  et~al., 2003, \mn@doi [\apjl] {10.1086/367929}, \href
  {http://adsabs.harvard.edu/abs/2003ApJ...583L..31G} {583, L31}

\bibitem[\protect\citeauthoryear{{Grevesse} \& {Sauval}}{{Grevesse} \&
  {Sauval}}{1998}]{Grevesse98}
{Grevesse} N.,  {Sauval} A.~J.,  1998, \mn@doi [\ssr]
  {10.1023/A:1005161325181}, \href
  {http://adsabs.harvard.edu/abs/1998SSRv...85..161G} {85, 161}

\bibitem[\protect\citeauthoryear{{Hu}, {Tout}, {Glebbeek}  \& {Dupret}}{{Hu}
  et~al.}{2011}]{Hu11}
{Hu} H.,  {Tout} C.~A.,  {Glebbeek} E.,   {Dupret} M.-A.,  2011, \mn@doi
  [\mnras] {10.1111/j.1365-2966.2011.19482.x}, \href
  {http://adsabs.harvard.edu/abs/2011MNRAS.418..195H} {418, 195}

\bibitem[\protect\citeauthoryear{{Istrate}, {Tauris}, {Langer}  \&
  {Antoniadis}}{{Istrate} et~al.}{2014}]{Istrate14}
{Istrate} A.~G.,  {Tauris} T.~M.,  {Langer} N.,   {Antoniadis} J.,  2014,
  \mn@doi [\aap] {10.1051/0004-6361/201424681}, \href
  {https://ui.adsabs.harvard.edu/abs/2014A&A...571L...3I} {571, L3}

\bibitem[\protect\citeauthoryear{{Istrate}, {Fontaine}  \& {Heuser}}{{Istrate}
  et~al.}{2017}]{Istrate17}
{Istrate} A.~G.,  {Fontaine} G.,   {Heuser} C.,  2017, \mn@doi [\apj]
  {10.3847/1538-4357/aa8958}, \href
  {https://ui.adsabs.harvard.edu/abs/2017ApJ...847..130I} {847, 130}

\bibitem[\protect\citeauthoryear{{Ivanova} et~al.,}{{Ivanova}
  et~al.}{2013}]{IvanovaRev13}
{Ivanova} N.,  et~al., 2013, \mn@doi [\aapr] {10.1007/s00159-013-0059-2}, \href
  {http://adsabs.harvard.edu/abs/2013A%26ARv..21...59I} {21, 59}

\bibitem[\protect\citeauthoryear{{Jeffery} \& {Saio}}{{Jeffery} \&
  {Saio}}{2013}]{JefferySaio13}
{Jeffery} C.~S.,  {Saio} H.,  2013, \mn@doi [\mnras] {10.1093/mnras/stt1360},
  \href {https://ui.adsabs.harvard.edu/abs/2013MNRAS.435..885J} {435, 885}

\bibitem[\protect\citeauthoryear{{Jeffery} \& {Saio}}{{Jeffery} \&
  {Saio}}{2016}]{JefferySaio16}
{Jeffery} C.~S.,  {Saio} H.,  2016, \mn@doi [\mnras] {10.1093/mnras/stw388},
  \href {http://adsabs.harvard.edu/abs/2016MNRAS.458.1352J} {458, 1352}

\bibitem[\protect\citeauthoryear{{Justham}, {Wolf}, {Podsiadlowski}  \&
  {Han}}{{Justham} et~al.}{2009}]{Justham09}
{Justham} S.,  {Wolf} C.,  {Podsiadlowski} P.,   {Han} Z.,  2009, \mn@doi
  [\aap] {10.1051/0004-6361:200810106}, \href
  {https://ui.adsabs.harvard.edu/abs/2009A&A...493.1081J} {493, 1081}

\bibitem[\protect\citeauthoryear{{Kilic}, {Stanek}  \& {Pinsonneault}}{{Kilic}
  et~al.}{2007}]{Kilic07}
{Kilic} M.,  {Stanek} K.~Z.,   {Pinsonneault} M.~H.,  2007, \mn@doi [\apj]
  {10.1086/522228}, \href
  {https://ui.adsabs.harvard.edu/abs/2007ApJ...671..761K} {671, 761}

\bibitem[\protect\citeauthoryear{{Kilkenny}, {Koen}, {O'Donoghue}  \&
  {Stobie}}{{Kilkenny} et~al.}{1997}]{Kilkenny97}
{Kilkenny} D.,  {Koen} C.,  {O'Donoghue} D.,   {Stobie} R.~S.,  1997, \mnras,
  \href {http://adsabs.harvard.edu/abs/1997MNRAS.285..640K} {285, 640}

\bibitem[\protect\citeauthoryear{{Kupfer} et~al.,}{{Kupfer}
  et~al.}{2019}]{Kupfer19}
{Kupfer} T.,  et~al., 2019, \mn@doi [\apjl] {10.3847/2041-8213/ab263c}, \href
  {https://ui.adsabs.harvard.edu/abs/2019ApJ...878L..35K} {878, L35}

\bibitem[\protect\citeauthoryear{{Landolt}}{{Landolt}}{1968}]{Landolt68}
{Landolt} A.~U.,  1968, \mn@doi [\apj] {10.1086/149645}, \href
  {https://ui.adsabs.harvard.edu/abs/1968ApJ...153..151L} {153, 151}

\bibitem[\protect\citeauthoryear{{Maxted} et~al.,}{{Maxted}
  et~al.}{2014}]{Maxted14}
{Maxted} P.~F.~L.,  et~al., 2014, \mn@doi [\mnras] {10.1093/mnras/stt2007},
  \href {https://ui.adsabs.harvard.edu/abs/2014MNRAS.437.1681M} {437, 1681}

\bibitem[\protect\citeauthoryear{{Michaud}, {Richer}  \& {Richard}}{{Michaud}
  et~al.}{2011}]{Michaud11}
{Michaud} G.,  {Richer} J.,   {Richard} O.,  2011, \mn@doi [\aap]
  {10.1051/0004-6361/201015997}, \href
  {http://adsabs.harvard.edu/abs/2011A%26A...529A..60M} {529, A60}

\bibitem[\protect\citeauthoryear{{Miller Bertolami}, {Althaus}, {Unglaub}  \&
  {Weiss}}{{Miller Bertolami} et~al.}{2008}]{MillerBertolami08}
{Miller Bertolami} M.~M.,  {Althaus} L.~G.,  {Unglaub} K.,   {Weiss} A.,  2008,
  \aap, 491, 253

\bibitem[\protect\citeauthoryear{{Opacity Project Team}}{{Opacity Project
  Team}}{1995}]{OP1}
{Opacity Project Team} 1995, {The Opacity Project, Vol. 1}.
Institute of Physics

\bibitem[\protect\citeauthoryear{{Opacity Project Team}}{{Opacity Project
  Team}}{1997}]{OP2}
{Opacity Project Team} 1997, {The Opacity Project, Vol. 2}.
Institute of Physics

\bibitem[\protect\citeauthoryear{{{\O}stensen} et~al.,}{{{\O}stensen}
  et~al.}{2010}]{Ostensen10}
{{\O}stensen} R.~H.,  et~al., 2010, \mn@doi [\aap]
  {10.1051/0004-6361/200913480}, \href
  {https://ui.adsabs.harvard.edu/abs/2010A&A...513A...6O} {513, A6}

\bibitem[\protect\citeauthoryear{{Ostlie} \& {Cox}}{{Ostlie} \&
  {Cox}}{1986}]{Ostlie86}
{Ostlie} D.~A.,  {Cox} A.~N.,  1986, \mn@doi [\apj] {10.1086/164824}, \href
  {https://ui.adsabs.harvard.edu/abs/1986ApJ...311..864O} {311, 864}

\bibitem[\protect\citeauthoryear{{Paxton}, {Bildsten}, {Dotter}, {Herwig},
  {Lesaffre}  \& {Timmes}}{{Paxton} et~al.}{2011}]{Paxton11}
{Paxton} B.,  {Bildsten} L.,  {Dotter} A.,  {Herwig} F.,  {Lesaffre} P.,
  {Timmes} F.,  2011, \mn@doi [\apjs] {10.1088/0067-0049/192/1/3}, \href
  {http://adsabs.harvard.edu/abs/2011ApJS..192....3P} {192, 3}

\bibitem[\protect\citeauthoryear{{Paxton} et~al.,}{{Paxton}
  et~al.}{2013}]{Paxton13}
{Paxton} B.,  et~al., 2013, \mn@doi [\apjs] {10.1088/0067-0049/208/1/4}, \href
  {http://adsabs.harvard.edu/abs/2013ApJS..208....4P} {208, 4}

\bibitem[\protect\citeauthoryear{{Paxton} et~al.,}{{Paxton}
  et~al.}{2015}]{Paxton15}
{Paxton} B.,  et~al., 2015, \mn@doi [\apjs] {10.1088/0067-0049/220/1/15}, \href
  {http://adsabs.harvard.edu/abs/2015ApJS..220...15P} {220, 15}

\bibitem[\protect\citeauthoryear{{Paxton} et~al.,}{{Paxton}
  et~al.}{2018}]{Paxton18}
{Paxton} B.,  et~al., 2018, \mn@doi [\apjs] {10.3847/1538-4365/aaa5a8}, \href
  {http://adsabs.harvard.edu/abs/2018ApJS..234...34P} {234, 34}

\bibitem[\protect\citeauthoryear{{Paxton} et~al.,}{{Paxton}
  et~al.}{2019}]{Paxton19}
{Paxton} B.,  et~al., 2019, \mn@doi [\apjs] {10.3847/1538-4365/ab2241}, \href
  {https://ui.adsabs.harvard.edu/abs/2019ApJS..243...10P} {243, 10}

\bibitem[\protect\citeauthoryear{{Pietrukowicz} et~al.,}{{Pietrukowicz}
  et~al.}{2017}]{Pietrukowicz17}
{Pietrukowicz} P.,  et~al., 2017, \mn@doi [Nature Astronomy]
  {10.1038/s41550-017-0166}, \href
  {http://adsabs.harvard.edu/abs/2017NatAs...1E.166P} {1, 0166}

\bibitem[\protect\citeauthoryear{{Ramsay}}{{Ramsay}}{2018}]{Ramsay18}
{Ramsay} G.,  2018, \mn@doi [\aap] {10.1051/0004-6361/201834604}, \href
  {https://ui.adsabs.harvard.edu/abs/2018A&A...620L...9R} {620, L9}

\bibitem[\protect\citeauthoryear{{Romero}, {C{\'o}rsico}, {Althaus}, {Pelisoli}
   \& {Kepler}}{{Romero} et~al.}{2018}]{Romero18}
{Romero} A.~D.,  {C{\'o}rsico} A.~H.,  {Althaus} L.~G.,  {Pelisoli} I.,
  {Kepler} S.~O.,  2018, \mn@doi [\mnras] {10.1093/mnrasl/sly051}, \href
  {http://adsabs.harvard.edu/abs/2018MNRAS.477L..30R} {477, L30}

\bibitem[\protect\citeauthoryear{{Shibahashi}}{{Shibahashi}}{2005}]{Shibahashi05}
{Shibahashi} H.,  2005, in {Alecian} G.,  {Richard} O.,   {Vauclair} S.,  eds,
  EAS Publications Series Vol. 17, EAS Publications Series. pp 143--148,
  \mn@doi{10.1051/eas:2005108}

\bibitem[\protect\citeauthoryear{{Stancliffe}, {Fossati}, {Passy}  \&
  {Schneider}}{{Stancliffe} et~al.}{2016}]{Stancliffe16}
{Stancliffe} R.~J.,  {Fossati} L.,  {Passy} J.-C.,   {Schneider} F.~R.~N.,
  2016, \mn@doi [\aap] {10.1051/0004-6361/201527099}, \href
  {http://adsabs.harvard.edu/abs/2016A%26A...586A.119S} {586, A119}

\bibitem[\protect\citeauthoryear{{Thoul}, {Bahcall}  \& {Loeb}}{{Thoul}
  et~al.}{1994}]{Thoul94}
{Thoul} A.~A.,  {Bahcall} J.~N.,   {Loeb} A.,  1994, \mn@doi [\apj]
  {10.1086/173695}, \href {http://adsabs.harvard.edu/abs/1994ApJ...421..828T}
  {421, 828}

\bibitem[\protect\citeauthoryear{{Townsend} \& {Teitler}}{{Townsend} \&
  {Teitler}}{2013}]{Townsend13}
{Townsend} R.~H.~D.,  {Teitler} S.~A.,  2013, \mn@doi [\mnras]
  {10.1093/mnras/stt1533}, \href
  {http://adsabs.harvard.edu/abs/2013MNRAS.435.3406T} {435, 3406}

\bibitem[\protect\citeauthoryear{{Wu} \& {Li}}{{Wu} \& {Li}}{2018}]{Wu18}
{Wu} T.,  {Li} Y.,  2018, \mn@doi [\mnras] {10.1093/mnras/sty1347}, \href
  {https://ui.adsabs.harvard.edu/abs/2018MNRAS.478.3871W} {478, 3871}

\makeatother
\end{thebibliography}

%\begin{thebibliography}{99}
%\end{thebibliography}

%%%%%%%%%%%%%%%%%%%%%%%%%%%%%%%%%%%%%%%%%%%%%%%%%%

%%%%%%%%%%%%%%%%% APPENDICES %%%%%%%%%%%%%%%%%%%%%


%% \section{Some extra material}


%%%%%%%%%%%%%%%%%%%%%%%%%%%%%%%%%%%%%%%%%%%%%%%%%%

% Don't change these lines
\bsp	% typesetting comment
\label{lastpage}
\end{document}